\documentclass[11pt]{article}
\usepackage{comment}
\usepackage{fancybox}
\usepackage{color}
\usepackage{titlesec}
\titlespacing*{\section}{0pt}{0.6ex}{0.6ex}
\titlespacing*{\subsection}{0pt}{0.6ex}{0.6ex}
\titlespacing*{\subsubsection}{0pt}{0.4ex}{0.4ex}

\titleformat{\section}
{\normalfont\fontsize{14}{24}\bfseries}{\thesection}{1em}{}
\titleformat{\subsection}
{\normalfont\fontsize{12}{24}\bfseries}{\thesubsection}{1em}{}

\titleformat{\subsubsection}
{\normalfont\fontsize{12}{24}\bfseries}{\thesubsubsection}{1em}{}
\definecolor{light-gray}{gray}{0.9}

\date{}
\usepackage[margin=1in]{geometry}
\usepackage{amsmath}
\usepackage{amssymb}
\usepackage{hyperref}
\usepackage{subcaption}
\usepackage{tikz}
\usepackage{pgfplots}

\DeclareMathOperator*{\argmin}{arg\,min}
\usepackage{amsthm}
\newtheorem{theorem}{Theorem}
\newtheorem{definition}[theorem]{Definition}

\title{ {\bf  Variational Encoders and Autoencoders : Information-theoretic Inference and Closed-form Solutions}}
\author{Karthik Duraisamy \\ {\em Department of Aerospace Engineering }\\ {\em University of Michigan, Ann Arbor}.}

\newcommand{\reals}{\mathbb{R}}

\DeclareMathOperator{\Tr}{\textrm{Tr}}
\newcommand{\norm}[1]{\| #1 \|}
\newcommand{\tr}{\textrm{Tr}}

\begin{document}
\maketitle

\begin{abstract}
This work develops  problem statements related to  encoders and autoencoders with the goal of elucidating variational formulations and establishing clear connections to  information-theoretic concepts.  Specifically, four problems with varying levels of input are considered : a)  The data, likelihood and prior distributions are given, b) The data and likelihood are given; c) The data and prior are given; d) the data and the dimensionality of the parameters is specified.  The first two problems seek encoders (or the posterior) and the latter two seek autoencoders (i.e. the posterior and the likelihood).  A variational Bayesian setting is  pursued, and detailed derivations are provided for the resulting optimization problem. Following this, a linear Gaussian setting is adopted, and closed form solutions are derived. Numerical experiments are also performed to verify expected behavior and assess convergence properties. Explicit connections are made to rate-distortion theory, information bottleneck theory, and the related concept of sufficiency of statistics is also explored. One of the  motivations of this  work is to present the theory and learning dynamics associated with variational inference and autoencoders, and to expose information theoretic concepts from  a computational science perspective. 
\end{abstract}
\vspace{1cm}

\tableofcontents

\vspace{1cm}

\section{Introduction}
Since being introduced by Kingma and Welling~\cite{kingma2013auto}  in 2014, variational autoencoders (VAEs) have become very popular in unsupervised learning and generative modeling. While there are excellent review articles~\cite{kingma2019introduction,doersch2016tutorial} and conference papers on this topic,  much of the attention in those articles (beyond the derivation of variational Bayes) is focused - and perhaps  rightly so - on the general setting of unsupervised learning, and on extensions of the formulation to address complex, real-world datasets. In preparing this work, the author has gained  useful insight from recent literature, but has found presentations therein to be  brief, and rapidly transitioned to complex problems.

In this work, simplified  problem statements are introduced, such that closed-form relationships can be derived where possible, and clear connections can be made to  information-theoretic concepts such as rate-distortion theory, information bottleneck, minimal sufficient statistic, etc. Further, in the author's own experience, practical implementations of VAEs suffer from many obfuscations (adhoc approximations, inadequate parametrization, sampling errors, convergence issues, etc.). To not risk falling behind the `veil' of a complex problem in which the results are not objectively quantifiable beyond prediction accuracy, there is merit in taking a simple problem, and verifying expected behavior. The author considers this to be a necessary step  before tackling more complex real-world problems (i.e. those problems for which VAEs are designed for).  Accordingly, a viewpoint of inference rather than learning is pursued.  Another important goal of this work is to elucidate concepts from information theory and connect them directly to variational inference,  again, benefitting from the prospect of closed-form solutions.

While  new problem statements, proofs,  connections and (potentially) new insight is brought to the fore, the author does not claim that this work presents {\em any} new {\em solutions} to {\em any of} the outstanding challenges in VAEs. That is the realm of NeurIPS, ICLR, ICML, etc.  We also do not address deep learning or neural networks in this work. There are excellent texts and resources in Information theory~\cite{cover2006elements,yeung2008information,mackay2003information} and the recent uptick in information-theoretic learning is a rich resource, though not  written for the mainstream computational science audience.
The main contribution of this work is to provide a principled set of problems and analytical solutions to help establish a better understanding of  approaches and algorithms for real problems. 

The organization of this manuscript is as follows: Section 2 introduces four encoding and autoencoding problems of interest to this work, and for the broader inference and learning communities; Section 3 introduces variational approaches to inference, and relevant concepts from information theory, including rate distortion theory, the idea of sufficient statistics, and information bottleneck. Section 4 maps the variational Bayesian approach to the encoder problems introduced in Section 2. Section 5 establishes the linear Gaussian setting for the present approach. Section 6 presents the analytical solution to variational encoder inference when the prior is specified, and a numerical verification is provided. The evolution of the numerical solution in the context of rate distortion theory and information bottleneck is documented. Section 7 presents the analytical solution to the encoder search problem when the prior is not known. Section 8 extends the analysis and numerics from Sections 6 and 7 to the Autoencoder case. A summary is provided in Section 9. The Appendix provides detailed derivations and proofs.

\section{A Quartet of Encoding and Autoencoding Problems}
This section will establish the notations followed in this paper, and formulate  encoding and autoencoding problems.

\noindent {\bf Notations:} 

Consider a continuous random variable $Y : \Omega \to \reals^{n}$, which will represent {\em data } or {\em observations}.  We will refer to realizations of  $Y$ as $y$, where $Y(\omega) = y : \omega \in \Omega $. The set of all possible realizations $y$ will be referred to as $\mathcal{Y}$. We will use the same upper case/lower case/symbol  notation for all random variables/realizations/set of realizations.  We define the probability density function $p(y) : \reals^n \to \reals^+$. 

Consider another continuous random variable $\Theta : \Omega  \to \reals^{m}$. We assume that the observations $y$ are generated by the {\em generative factors} $\theta$ via a density $p(\theta)$ and  a   {\em  model} or {\em likelihood}  $p(y|\theta)$. This  generates the  joint distribution $p(\theta,y) = p(\theta) p(y|\theta)$, and the `data distribution' $p(y) = \int_\Theta p(\theta,y) d\theta$. The essence of this work is to study  techniques to extract approximations to $p(\theta|y)$ and $p(y|\theta)$ given varying levels of information, as detailed below. These approximations will be denoted by $p_\phi(\theta|y)$ and $p_\psi(y|\theta)$, where $\phi,\psi$ denote the parameters describing the probability density functions (PDFs).   It is implicit in the rest of the manuscript that all covariance matrices are symmetric and positive semi-definite, and thus we will not explicitly state this in the optimization problem statements.

Finally, we acknowledge slight abuse of notation when specifying Gaussian probability density functions. For instance, while specifiying the likelihood, in contrast to the conventional $Y|\Theta \sim \mathcal{N}(A \theta,S)$ and the associated  PDF $p_{Y|\Theta}(Y=y|\Theta=\theta)$, the  notation $p(y|\theta) = \mathcal{N}(y;A \theta,S)$ is used to unclutter the presentation (and avoid multiple subscripts). With this notation, it is easier to distinguish $p(y|\theta)$ and $p_\psi(y|\theta)$.

We now introduce four problems of interest to this work:

\noindent {\bf Encoder Inference: }

Given $p(y) , p(y|\theta), p(\theta) $, the goal of the encoder inference problem is to extract the encoder $p(\theta|Y(\omega)=y)  \  \  \forall y \in \mathcal{Y}$. Clearly, for a given realization of the data, one can use the Bayes posterior on the parameters
\begin{equation}
p(\theta|y)  = \frac{p(y|\theta)p(\theta)}{p(y)}.
\label{eq:true_encoder}
\end{equation}

In practice, however,  there are two challenges:

$\bullet$ The Bayesian inference solution as stated above may be intractable in high dimensions

$\bullet$ Even if the inference problem above is tractable, the stated goal  is to not just extract an encoder for a given $Y(\omega)=y$, but rather to extract an encoder $\forall y \in \mathcal{Y}$. 

Thus, the goal is to extract an approximate encoder, which we will refer to as $p_\phi(\theta|y)$, where the $\phi$ denotes a parametrization. It is notable that the approximate encoder will induce a new joint distribution $p_\phi(\theta,y) = p(y) p_\phi(\theta|y)$ and a new marginal distribution $p_\phi(\theta) = \int_Y p_\phi(\theta,y) dy$.

\noindent {\bf Encoder Search: }

In this case, we are only given $p(y)$ and $p(y|\theta)$, and the goal is to extract the encoder $p_\phi(\theta|y)  \  \  \forall y \in \mathcal{Y}$. However, a key difference from the encoder inference problem is that $p(\theta)$ is not given. Therefore, from a Bayesian standpont, there is a need to not just efficiently determine the posterior $p_\phi(\theta|y)$ as in the encoder inference  problem, but the prior (marginal) $p_\phi(\theta)$ also has to be chosen/extracted appropriately.

\begin{center}
	\fbox{\colorbox{light-gray}{
			\begin{minipage}[t]{0.95\textwidth}
				
				\begin{definition}[{\bf Encoder/Autoencoder Inference and Search  Problems} ]\label{def:problem}		
				\end{definition}

				Consider random variables   $Y : \Omega \to \reals^n$ and $\Theta : \Omega \to \reals^m$. 
				\vspace{0.25cm}	
				
				{\bf Encoder Inference}

				Given :  $p(y), p(y|\theta), p(\theta)$
				
				Required: $p_\phi(\theta|y)$, where $p_\phi(\theta,y) \triangleq p_\phi(\theta|y) p(y) $.
				
				{\bf Encoder Search}
				
				Given :  $p(y), p(y|\theta)$
				
				Required: $p_\phi(\theta|y)$, where $p_\phi(\theta,y) \triangleq p_\phi(\theta|y) p(y)$.

				{\bf Autoencoder Inference}

				Given :   $p(y),  p(\theta)$
				
				Required: $p_\phi(\theta|y), p_\psi(y|\theta)$, where $p_\phi(\theta,y) \triangleq p_\phi(\theta|y) p(y), \ \  p_\psi(\theta,y) \triangleq p_\psi(y|\theta) p(\theta)$.
				
				{\bf Autoencoder Search}

				Given :   $p(y),  m$ 
				
				Required: $p_\phi(\theta|y), p_\psi(y|\theta)$, where $p_\phi(\theta,y) \triangleq p_\phi(\theta|y) p(y), \ \  p_\psi(\theta,y) \triangleq p_\psi(y|\theta) p_\phi(\theta) $.
				
				\vspace{0.25cm}	
				Notes: 
				
				$\bullet$ $\phi,\psi$ represent parameters describing the associated probability densities. 
				
				$\bullet$ In all the cases,  the given / required densities are provided/sought $ \forall y \in \mathcal{Y}$
				
			\end{minipage}
		}
	}
\end{center}

\noindent {\bf Autoencoder Inference: }

In this case, we are only given $p(y)$ and $p(\theta)$, and the goal is to extract an approximate encoder $p_\phi(\theta|y)$ and  decoder $p_\psi(y|\theta) $. As in the encoder problems above, the encoder induces a joint distribution $p_\phi(\theta,y) = p(y) p_\phi(\theta|y)$ and  marginal distribution $p_\phi(\theta) = \int_Y p_\phi(\theta,y) dy$. In this case, however, the decoder induces another joint distribution $p_\psi(\theta,y) = p_\psi(y|\theta) p(\theta)$.

\noindent {\bf Autoencoder search: }

In this case, we are only given $p(y)$ and the dimension $m$ of the latent variable $\theta$. The goal is to extract an encoder $p_\phi(\theta|y)$ and a decoder $p_\psi(y|\theta) $ based on this input. Note that, in contrast to the Autoencoder inference, the second induced joint distribution  is  $p_\psi(\theta,y) = p_\psi(y|\theta) p_\phi(\theta)$.

These four problems  are concisely stated in definition~\ref{def:problem}.

\section{A Quick Tour of Variational Approaches for Inference}
In this section, a brief introduction will be given to relevant concepts in variational inference and information theory.

\begin{center}
	\fbox{\colorbox{light-gray}{
			\begin{minipage}[t]{1.01\textwidth}
				\begin{definition}[
					{\bf Relevant quantities from information theory: The discrete (Shannon) case}
					]\label{def:info}		
				\end{definition}

				Shannon's~\cite{shannon1948mathematical} information measure $i(.)$ satisfies the following axioms:
				
				$\bullet$ Information acquired about an event  should depend on the probability $P$ of that event.
				
				$\bullet$ $i(P)\ge 0$, $i(1)= 0$.
				
				$\bullet$ For two independent events $i(P_1 P_2) = i(P_1) + i(P_2)$.
				
				$\bullet$ $i(P)$ is continuous and monotonic in $P$.
				
				\noindent Shannon~\cite{shannon1948mathematical} showed that $i(P)=-log(P)$ uniquely satisfies these axioms.
				
				Inspired by the development in Berger~\cite{berger1971}, we consider a discrete sample space $\Omega$ and random variables $X(j)=j$ for $1 \leq j \leq m$ and $Y(k) = k$
				for $1 \leq k \leq n$. The associated probability distributions are $P_X$ and $P_Y$.  The table below presents a concise description of key quantities for this setup. All quantities are measured in bits.

				\vspace{2mm}\hspace{-7mm}\label{tab:info}
				{\small
					\begin{center}
						\begin{tabular}{|l|l|l|}
							\hline
							Quantity & Definition & Description  \\
							\hline \hline
							Self-Information &  $i(j)=-\log[P_X(j)]$ & Information acquired after observing $X=j$  \\ \hline
							Entropy &  $H(X)=-\Sigma_j P_X(j)\log[P_X(j)]$ & Average uncertainty associated with $X$ \\ \hline
							Conditional  &  $i(j|k)=-\log[P_{X|Y}(j|k)]$ & Information acquired upon observing $X=j$, \\
							Self-Information & &  given we know that  $Y=k$ has occurred.\\ \hline
							Local  &  $i(j;k)=i(j)-i(j|k)$ & Difference between self information	\\ 
							Mutual Information &   & and conditional self information $i(j;k) = i(k;j)$ 
							\\ \hline
							Conditional Entropy &  $H(X|Y)=$ & Average uncertainty associated with $X$ \\
							& $-\Sigma_{j,k} P_{XY}(j,k) \log[P_{X|Y}(j|k)]$ &  after Y has been observed \\ \hline
							Mutual Information &  $I(X;Y)=$ &  Average information that knowledge of   \\
							& $\Sigma_{j,k} P_{XY}(j,k) \log\left[\frac{P_{XY}(j,k)}{P_X(j)P_Y(k)}\right]$ & Y supplies about the observation of X   \\
							\hline
							
						\end{tabular}
					\end{center}
				}
				
				The communication theory interpretation of cross entropy of an estimated distribution $P_Y$ relative to a true distribution $P_X$ over the same
				set of underlying events is the number of bits required to encode $P_X$ using $P_Y$. 
				
				$$H(P_X,P_Y) = \Sigma_{j} P_X(j) \log\left[\frac{1}{P_Y(j)}\right].$$
				
				The KL divergence or relative entropy from a distribution $P_X$ to a distribution $P_Y$ measures how one probability distribution is different from a second reference probability distribution and is defined as
				$${KL}(P_X || P_Y) = \Sigma_{j} P_X(j) \log\left[\frac{P_X(j)}{P_Y(j)}\right] = H(P_X,P_Y) - H(X) .$$ 
				
				Appendix~\ref{sec:continuous} gives extensions to the continuous case, and provides explicit expressions for Gaussian distributions.
				
			\end{minipage}
		}
	}
\end{center}

We will not yet map these approaches to the problem  statements in the previous section, but will point out that  $y$ can be considered as observations, and the variational approaches seek modeled distributions  (of $p(y)$) in terms of latent variables $Z : \Omega \to \reals^k$. Latent variables are typically unobserved, and serve several purposes. For instance, in image processing $y$ may refer to an image and $z$ may refer to an image classifier. In reduced order modeling, $y$ may refer to realizations of the state variable and $z$ may refer to realizations of the reduced dimensional variable. 
The modeled distribution is defined as $p_\psi(y) = \int_Z p_\psi(y,z) dz$, where $\psi$ are parameters,  and $p_\psi(y,z) = p_\psi(y|z) p_\psi(z)$ is the  modeled joint distribution.  We then have $p_\psi(z|y) = p_\psi(y|z) p_\psi(z)/p_\psi(y)$, which represents an encoding of the observation in terms of the latent variables. However, $p_\psi(z|y)$ may be hard to compute for various reasons (which will be explained soon). Thus, we define an approximation $p_\phi(z|y)$. We will refer to $p_\phi(z|y)$ as the encoding distribution, and $p_\psi(y|z)$ as the decoding distribution.

\subsection{Variational Inference}
In the  general case, variational inference~\cite{wainwright2008introduction,blei2017variational, kingma2019introduction} seeks to minimize the distance between the data and the modeled distributions. Expressing the distance as a KL Divergence, we have:
\begin{align}
   KL[p(y)||p_\psi(y)] =  \mathbb{E}_Y [\log p(y) ] + \mathbb{E}_Y [-\log p_\psi(y)] .
\end{align}
The first term in the RHS  is the negative of the entropy of the data distribution (i.e. $ \mathbb{E}_Y [\log p(y) ] = -H(Y)$) and is of course independent of the model. The second term   $\mathbb{E}_Y [-\log p_\psi(y)]  = H(p(y),p_\psi(y))$ is the cross-entropy between the data and modeled distributions. Let's expand this by starting with the  Bayes rule
\begin{align*}
p_\psi(y) &=  \frac{p_\psi(y|z) p_\psi(z)}{p_\psi(z|y)}  =   \frac{p_\psi(y|z) p_\psi(z)}{p_\psi(z|y)}  \frac{p_\phi(z|y)}{p_\phi(z|y)}\\
-\log p_\psi(y) &=   -\log p_\psi(y|z) +\log \left[ \frac{ p_\phi(z|y)}{p_\psi(z)}  \right] - \log \left[ \frac{ p_\phi(z|y)}{p_\psi(z|y)} \right] \\
\end{align*}
Let's now take an expectation over $p_\phi(z,y) = p(y) p_\phi(z|y)$
\begin{align*}
-\mathbb{E}_Y \log p_\psi(y) &=  - \mathbb{E}_Y \mathbb{E}_{Z|Y}^\phi [\log p_\psi(y|z)]+ \mathbb{E}_Y \mathbb{E}_{Z|Y}^\phi \log \left[ \frac{ p_\phi(z|y)}{p_\psi(z)}  \right]  -  \mathbb{E}_Y [ KL (p_\phi(z|y) || p_\psi(z|y)) ] .
\end{align*}

The first two terms in the RHS constitute  the expectation of the (negative) Evidence lower bound (ELBO), which we define as 
\begin{align} \label{eq:ELBO}
	\mathcal{L}(\theta,\phi) \triangleq  - \mathbb{E}_Y \mathbb{E}_{Z|Y}^\phi [\log p_\psi(y|z)]+ \mathbb{E}_Y \mathbb{E}_{Z|Y}^\phi  \log \left[ \frac{ p_\phi(z|y)}{p_\psi(z)}  \right].
\end{align}

Thus, we have
\begin{align*}
KL[p(y)||p_\psi(y)] = -H(Y)   -  \mathbb{E}_Y [ KL (p_\phi(z|y) || p_\psi(z|y)) ]  + \mathcal{L}(\phi,\psi)
\end{align*}

or 

\begin{align}
\label{eq:VIgen}
 KL[p(y)||p_\psi(y)] + H(Y)   +  \mathbb{E}_Y [ KL (p_\phi(z|y) || p_\psi(z|y)) ]  =   \mathcal{L}(\phi,\psi).
\end{align}

Given $H(Y)$ is a constant, minimizing $\mathcal{L}(\phi,\psi)$ is equivalent to minimizing $KL[p(x)||p_\psi(x)] +  \mathbb{E}_Y [ KL (p_\phi(z|y) || p_\psi(z|y)) ]$, effectively driving the modeled distribution to match the data distribution.

Note: Eq.~\ref{eq:ELBO} can also be written in the form of an energetic term and an entropic term
\begin{align}
\mathcal{L}(\theta,\phi) \triangleq  - \mathbb{E}_Y \mathbb{E}_{Z|Y}^\phi  \left[\log p_\psi(y,z) \right] + \mathbb{E}_Y \mathbb{E}_{Z|Y}^\phi  \log \left[ p_\phi(z|y)  \right].
\end{align}

\begin{center}
	\fbox{\colorbox{light-gray}{
			\begin{minipage}[t]{1.01\textwidth}
				
				\begin{definition}[
					{\bf Relevant inequalities from information theory}
					]\label{def:inequalities}		
				\end{definition}

				The following useful inequalities hold true for a pair of random variables $X$ and $Y$ :

               			$$ I(X;Y) \geq 0 $$
				
				$$H(X,Y) \leq H(X) + H(Y).$$
				
				$$I(X;Y)=H(X)-H(X|Y) =H(Y)-H(Y|X) =H(X)+H(Y)-H(X,Y) = I(Y;X)$$
				
				$$H(X|Y) \leq H(X)$$
				
				Gibbs inequality : ${KL}(P_X||P_Y) \geq 0$ ; ${KL}(P_Y||P_X) \geq 0$
				
				$$I(X;Y) = KL(P_{XY} || P_X P_Y) = \mathbb{E}_Y[KL(P_{X|Y} || P_X)].$$

				If $f,g$ are bijective mappings, then $I(f(X);g(Y)) = I(X;Y)$.

			The following relationships are only true in the discrete case:	$$H(X) \geq 0 \ \ ; \ \ H(Y) \geq 0 \ \ ; \  \ I(X;Y) \leq H(X) \ \ ; \ \ I(X;Y) \leq H(Y).$$ Also refer appendix ~\ref{sec:continuous}.

			\end{minipage}
		}
	}
\end{center}

\subsection{Rate Distortion}
Consider  random variables $Z : \Omega  \to \reals^{k}$  and  $Y : \Omega \to \reals^{n}$ which can be assumed to represent latent variables, and observations, respectively. Given  the distribution $p(y)$, a model $\tilde{y}(z)$, where $\tilde{y} :  \reals^k \to \reals^n$,  a distortion metric $d(y,\tilde{y}) : \reals^{n} \times \reals^{n} \to \reals$, and a bound $D \in \reals$,  the rate distortion problem~\cite{berger1971} seeks an encoder $p(z|y)$ in the following from
\begin{equation}
R(D) = \min_{p(z | y)} I(Z; Y)   \ \  \textrm{subject to} \ \ 
\mathbb{E}_{Z,Y}\left[d(y, \tilde{y}(z))\right] \leq D.
\end{equation}

It has been shown~\cite{cover2006elements} that $R(D)$ is a monotonically decreasing function. 
Writing the Lagrange function of the RD problem, we have
\begin{equation}\label{rd:lag}
\min_{p(z | y)}  \mathcal{J}(\beta)= \min_{p(z | y)} I(Z; Y) + \beta(\mathbb{E}_{Z,Y}\left[d(y, \tilde{y}(z))\right] -D).
\end{equation}

The Blahut Arimoto agorithm~\cite{blahut1972computation,arimoto1972algorithm} is an alternating minimization algorithm to minimize the Lagrangian in Eqn.~\ref{rd:lag} and the solution is presented as a fixed point iteration
\begin{equation}\label{eq:ba}
p(z|y) = \frac{p(z)e^{-\beta d(y, \tilde{y}(z))}}
{\int_Z p(z)e^{-\beta d(y, \tilde{y}(z))} dz}   \  \ ; \  \ p(z) = \int_Y p(y) p(z|y) dy .
\end{equation}

\subsection{Minimal Sufficient Statistic and Information Bottleneck}
\label{subsec:IB}
Consider random variables   $Z : \Omega  \to \reals^{k}$  ,   $Y : \Omega \to \reals^{n}$ and  $X : \Omega \to \reals^{m}$. In the present context, we can consider a Markov chain $X \to Y \to Z$ and assume those variables to represent relevant, observed and latent variables, respectively.

$Z$ is a minimal sufficient statistic~\cite{shamir2010learning} for $X$ if it satisfies 

\begin{equation}
\min_Z   I(Y;Z)  \  \ \textrm{subject to} \ \ I(X;Z) = I(X;Y).
\end{equation}

 Such a $Z$ contains all the information about $X$ while retaining the minimum possible information about $Y$ ~\cite{hafez2019information}.

 Given the above optimization problem is  intractable in practical problems, the information bottleneck~\cite{tishby2000information} seeks an encoder  given the joint distribution $p(y,x)$, and  $\beta \in \reals^+$ in the following form
\begin{equation}\label{ib:lag}
\min_{p(z | y)}  \mathcal{I}(\beta)= \min_{p(z | y)} I(Z; Y) - \beta I(Z; X).
\end{equation}

The Blahut Arimoto-type agorithm  to minimize the Lagrangian in Eqn.~\ref{ib:lag} was given by Tishby et al.~\cite{tishby2000information}
\begin{equation}\label{eq:ba_IB}
p(z|y) = \frac{p(z)e^{-\beta KL[p(x|y)||p(x|z)]}}
{\int_Z p(z)e^{-\beta KL[p(x|y)||p(x|z)])} dz}   \  \ ; \  \ p(z) = \int_Y p(y) p(z|y) dy   \ \ ;  \  \ p(x|z) =\frac{1}{p(z)} \int_Y p(x,y) p(z|y) dy 
\end{equation}

Note: The data processing inequality yields  $I(X;Y) \geq I(X;Z)$. Therefore, instead of the MSS, one can pose

\begin{equation}
\min_Z   I(Y;Z)  \  \ \textrm{subject to} \ \  I(X;Y) - I(X;Z) > q,
\end{equation}
where $q \geq 0$ is some threshold.
The Lagrangian can be written as 
\begin{equation}
\min_Z   I(Y;Z)  + \beta  (I(X;Y) - I(X;Z) - q),
\end{equation}
which in turn is equivalent to 
\begin{equation}
\min_Z   I(Y;Z)  - \beta   I(X;Z).
\end{equation}

Tishby et al.~\cite{tishby2000information} also point out that Information bottleneck is equivalent to Rate distortion problems with a distortion $ d(y,\tilde{y}(z)) = KL[p(x|y)||p(x|z)]$.

\begin{center}
	\fbox{\colorbox{light-gray}{
			\begin{minipage}[t]{0.95\textwidth}
				
				\begin{definition}[{\bf Variational Statements of Encoder/Autoencoder Inference and  Search} ]\label{def:solution}		
				\end{definition}
								Consider random variables   $Y : \Omega \to \reals^n$ and $\Theta : \Omega \to \reals^m$. In the variational problems below, we seek $\hat{p}_\phi(\theta|y)$ (and where applicable, $\hat{p}_\psi(y|\theta)$) as probability density functions.
				
				{\bf Variational Encoder Inference (VEI)}

				Given :  $p(y), p(y|\theta), p(\theta)$
				
				Required: 
				\begin{equation}
				\hat{p}_\phi(\theta|y) = \argmin_{p_\phi(\theta|y)}  \ \  \mathbb{E}_{\Theta,Y}^\phi \log \left[\frac{p_\phi(\theta|y)}{p(\theta)} \right] +  \mathbb{E}_{\Theta,Y}^\phi [-\log p(y|\theta)] 
				\end{equation}
				
				{\bf Variational Encoder Search (VES)}

				Given :   $p(y), p(y|\theta)$
				
				Required: 	\begin{equation}
				\hat{p}_\phi(\theta|y) = 	\argmin_{p_\phi(\theta|y)}  \ \ \mathbb{E}_{\Theta,Y}^\phi \log \left[ \frac{p_\phi(\theta|y)}{p_\phi(\theta)} \right] + \mathbb{E}_{\Theta,Y}^\phi [-\log p(y|\theta)]  .
				\end{equation}
				
					{\bf Variational Autoencoder Inference (VAEI)}

				Given :  $p(y), p(\theta)$
				
				Required: 
				\begin{equation}
				\hat{p}_\phi(\theta|y),\hat{p}_\psi(y|\theta)   = \argmin_{p_\phi(\theta|y),p_\psi(y|\theta)}  \ \  \mathbb{E}_{\Theta,Y}^\phi \log \left[\frac{p_\phi(\theta|y)}{p(\theta)} \right] +  \mathbb{E}_{\Theta,Y}^\phi [-\log p_\psi(y|\theta)] 
				\end{equation}
				
				{\bf Variational Autoencoder Search (VAES)}

				Given :   $p(y)$
				
				Required: 	\begin{equation}
				\hat{p}_\phi(\theta|y),\hat{p}_\psi(y|\theta) = 	\argmin_{p_\phi(\theta|y),p_\psi(y|\theta)}  \ \ \mathbb{E}_{\Theta,Y}^\phi \log \left[ \frac{p_\phi(\theta|y)}{p_\phi(\theta)} \right] + \mathbb{E}_{\Theta,Y}^\phi [-\log p_\psi(y|\theta)]  .
				\end{equation}

				$\bullet$  Note: 	$\phi, \psi$ represent parameters of the modeled encoder and decoder, respectively.  $E_{\Theta,Y}^\phi  [\cdot] = \int_Y \int _\Theta [\cdot] p_\phi(\theta,y) d \theta dy $ and	$ p_\phi(\theta) = \int_Y [ p_\phi(\theta|y) ] p(y) dy $.

					As a related problem we also define the
				
				{\bf $\beta-$Variational Encoder Search ($\beta-$VES)}

				Given :   $p(y), p(y|\theta)$
				
				Required: 	\begin{equation}
				\hat{p}_\phi(\theta|y) = 	\argmin_{p_\phi(\theta|y)}  \ \ \mathbb{E}_{\Theta,Y}^\phi \log \left[ \frac{p_\phi(\theta|y)}{p_\phi(\theta)} \right] + \beta \mathbb{E}_{\Theta,Y}^\phi [-\log p(y|\theta)],
				\end{equation}
				where $\beta \in \reals^+$.

			\end{minipage}
		}
	}
\end{center}

\section{Mapping Variational Bayes to the Encoding Problems}

In this section, we map the general variational inference approach to the encoder inference and search  problems in a Bayesian setting.

\subsection{Variational Encoder Inference}
In variational encoder inference, we are given $p(y), p(y|\theta), p(\theta)$. In
 Eq.~\ref{eq:VIgen},  $p_\psi(y|z) = p(y|\theta)$ and $p_\psi(z)= p(\theta)$, and  $p_\psi(y) = p(y)$. Therefore
 
\begin{align*}
H(Y)  &=  - \mathbb{E}_Y \mathbb{E}_{\Theta|Y}^\phi [\log p(y|\theta)] + \mathbb{E}_Y \mathbb{E}_{\Theta|Y}^\phi \log \left[\frac{p_\phi(\theta|y)}{p(\theta)} \right]  -  \mathbb{E}_Y [ KL (p_\phi(\theta|y) || p(\theta|y)) ]\\
 &=  - \mathbb{E}_Y \mathbb{E}_{\Theta|Y}^\phi [\log p(y|\theta)] + \mathbb{E}_Y \mathbb{E}_{\Theta|Y}^\phi \log \left[\frac{p_\phi(\theta|y)}{p_\phi(\theta)} \right] +  \mathbb{E}_Y \mathbb{E}_{\Theta|Y}^\phi \log \left[\frac{p_\phi(\theta)}{p(\theta)} \right] -  \mathbb{E}_Y [ KL (p_\phi(\theta|y) || p(\theta|y)) ]\\
 &=  - \mathbb{E}_Y \mathbb{E}_{\Theta|Y}^\phi [\log p(y|\theta)] + \mathbb{E}_Y \mathbb{E}_{\Theta|Y}^\phi \log \left[\frac{p_\phi(\theta|y)}{p_\phi(\theta)} \right] +  KL \left[p_\phi(\theta) || p(\theta) \right] -  \mathbb{E}_Y [ KL (p_\phi(\theta|y) || p(\theta|y)) ]
\end{align*}

The LHS term is the entropy of the data.

On the RHS, the first term is called the reconstruction loss $L_{rec}$ and is an approximation to $H(Y|\Theta)$

The second term is the mutual information $I_\phi(Y; \Theta)$ between the variables $Y,\Theta$ with respect to the induced distribution $p_\phi(\theta,y)$ and is an approximation to  $I(Y ; \Theta)$.

The third term, which we will refer to as $T_\phi$ is a consequence of the fact that the induced marginal distribution $p_\theta(\theta)$ (this is referred to as the  aggregated posterior in the ML community) is different from the true parameter distribution $p(\theta)$.

Typically, the second and third terms are combined and referred to as the regularization loss $L_{reg} = I_\phi(Y; \Theta) + T_\phi  $.

The fourth term is the residual $D_\phi \leq 0 $ and quantifies the distance between the approximate encoder and the true encoder over the entire data distribution.

Thus, we have

\begin{equation}
H(Y) =L_{rec}+ I_\phi(Y;\Theta) + T_\phi + D_\phi =  L_{rec} + L_{reg} + D_\phi = \mathcal{L}_{VEI}(\phi) +  D_\phi.
\label{eq:VI}
\end{equation}

 Since $H(Y)$ is a constant, minimizing $\mathcal{L}_{VEI}(\phi) \triangleq L_{rec} + L_{reg}$ with respect to $\phi$ is  equivalent to minimizing $D_\phi$, which means we minimize the average (over the data) distance between the approximate encoder and the true encoder.
 
 \subsection{Variational Encoder Search}
 The development for variational encoder search is similar. The key difference is that $p(\theta)$ is not given, and thus the regularization part of the loss function is just $ I_\phi(Y;\Theta)$ . Thus, minimizing $\mathcal{L}_{VES}(\phi) \triangleq L_{rec} + I_\phi(\Theta;Y)$ is equivalent to minimizing $\mathbb{E}_Y[KL(p_\phi(\theta|y)||p(\theta|y))] - KL(p_\phi(\theta)||p(\theta))$. We will examine connections with  Rate Distortion theory in the following sections.

\section{A Linear Gaussian Setting}

We now consider a linear Gaussian setting for the above problems by considering the generating distribution $p(\theta)  \triangleq \mathcal{N}(\theta; \mu_\Theta,\Sigma_\Theta)$, and $p(y|\theta)  =  \mathcal{N}(y; A \theta, S)$, where $A \in \reals^{n \times m}$, $\mu_\Theta \in \reals^m$, $\Sigma_\Theta  \in \reals^{m \times m}$, $S \in \reals^{m \times m}$, are all given and constant.

The joint distribution of the data $Y$ and the parameter $\Theta$ is then
\begin{equation}
p(\theta, y)  =   \mathcal{N}\left(\left[ \begin{array}{c} \theta \\ y \end{array} \right];\left[ \begin{array}{c} \mu_\Theta \\ A \mu_\Theta \end{array} \right],
\left[
\begin{array}{cc}
\Sigma_{\Theta} &   \Sigma_\Theta A^T \\
A \Sigma_{\Theta} &  A \Sigma_\Theta A^T + S
\end{array}
\right]\right).
\end{equation}

Also,
\begin{align}
p(\theta|y) & =  \mathcal{N}(\theta ; R y + \mu_\Theta-R A \mu_\Theta , \Sigma_\Theta - R A \Sigma_\Theta), \\
\textrm{where} \  \  R &= \Sigma_\Theta A^T (A \Sigma_\Theta A^T + S)^{-1}.
\end{align}

For notational simplicity, we will define  $p(\theta|y)  =  \mathcal{N}(\theta;R y + b, Q)$, where $b \triangleq \mu_\Theta-R A \mu_\Theta = \mu_\Theta - R \mu_Y$  and $Q \triangleq  \Sigma_\Theta - R A \Sigma_\Theta$.

Alternately, we can write

\begin{equation} \label{eq:generating}
R=  Q A^T S^{-1} \  \ ; \  \ b=  Q \Sigma_\Theta^{-1} \mu_\Theta \ \ ; \ \ Q = \left(A^T S^{-1} A + \Sigma_\Theta^{-1}\right)^{-1}.\end{equation}

Consequently,

\begin{align}
H(Y) &= \frac{n}{2}\log(2 \pi e) + \frac{1}{2}\log|\Sigma_Y|  \\ 
I(Y;\Theta) &= \frac{1}{2} \log \left[  \frac{| \Sigma_Y|}{|S|} \right] =   \frac{1}{2} \log \left[  \frac{|\Sigma_\Theta|}{|Q|} \right] \\ 
H(Y|\Theta) &=  \frac{n}{2}\log(2 \pi e) + \frac{1}{2} \log |S|.
\end{align}

\subsection{Variational Encoder and Information Budgets}
In the encoding process, we are given the true parameter distribution $p(\theta)$ and the data distribution $p(y)$ and the likelihood  $p(y|\theta)$. We define an encoder $p_\phi(\theta|y)  = \mathcal{N}(\theta; R_\phi y + b_\phi, Q_\phi)$, where $R_\phi \in \reals^{m \times n}, b_\phi \in \reals^{m \times 1}, Q_\phi \in \reals^{m \times m}$. The induced joint distribution is thus
\begin{equation}
p_\phi(\theta, y)  =   \mathcal{N}\left(\left[ \begin{array}{c} \theta \\ y \end{array} \right] ; \left[ \begin{array}{c} R_\phi \mu_Y+ b_\phi \\ \mu_Y \end{array} \right],
\left[
\begin{array}{cc}
R_\phi \Sigma_Y R_\phi^T + Q_\phi &  R_\phi  \Sigma_Y \\
\Sigma_{Y} R_\phi^T &  \Sigma_Y.
\end{array}
\right]\right).
\end{equation}

Define $\mu_\Theta^\phi \triangleq R_\phi \mu_Y + b_\phi$ and $\Sigma_\Theta^\phi \triangleq R_\phi \Sigma_Y R_\phi^T + Q_\phi$.

It is notable that 
\begin{align*}
p_\phi(y|\theta) &  = \mathcal{N}(y;\mu_Y+\Sigma_Y R_\phi^T \Sigma_\Theta^{\phi -1} (\theta - \mu_\Theta^\phi),\Sigma_Y-\Sigma_Y R_\phi^T \Sigma_\Theta^{\phi -1}R_\phi \Sigma_Y) \\
&  =  \mathcal{N}(y;(R_\phi^T Q_\phi^{-1} R_\phi + \Sigma_Y^{-1})^{-1} R_\phi^T Q_\phi^{-1}  \theta+(R_\phi^T Q_\phi^{-1} R_\phi + \Sigma_Y^{-1})^{-1}( \Sigma_Y^{-1} \mu_Y-R_\phi^T Q_\phi^{-1} b_\phi),(R_\phi^T Q_\phi^{-1} R_\phi + \Sigma_Y^{-1})^{-1} ) \\
\end{align*}.

Also,
\begin{equation*}
H(Y) = \frac{n}{2}\log(2 \pi e) + \frac{1}{2}\log|\Sigma_Y|   \ \ ; \ \ 
I_\phi(Y;\Theta) =\frac{1}{2} \log \left[  \frac{|\Sigma_\Theta^\phi|}{|Q_\phi|} \right].
\end{equation*}

Given these definitions, the relevant terms in Eqn.~\ref{eq:VI} are provided in table ~\ref{tab:full}. The derivation is provided in  Appendix ~\ref{sec:TableProof}.

\begin{table}[htbp]
	{   
		\caption{Full Space View}
		\label{tab:full}
		\begin{center}
			\begin{tabular}{|c|c|l|} \hline
				\bf Term & \bf Notation  & \bf Expression  \\  \hline \hline
			 	$- \mathbb{E}_Y \mathbb{E}_{\Theta|Y}^\phi [\log p(y|\theta)] $ & $L_{rec}$   &   $\frac{n}{2}\log{2 \pi} + \frac{1}{2} \log|S| + \frac{1}{2}(\mu_Y - A \mu_\Theta^\phi)^T S^{-1} (\mu_Y-A \mu_\Theta^\phi)$ \\
			 	&& $+   \frac{1}{2}  \tr\left[S^{-1} \Sigma_Y  - 2 S^{-1} A  R_\phi \Sigma_Y  +  S^{-1} A \Sigma_\Theta^\phi A^T  \right] $ \\ \hline
			 	$\mathbb{E}_Y [ KL (p_\phi(\theta|y) || p(\theta)) ]$ & $L_{reg}$ & $\frac{1}{2}  \log \left[\frac{|{\Sigma}_\Theta|}{|{Q}_\phi|}\right]  + \frac{1}{2} \tr[\Sigma_\Theta^{-1} {\Sigma}^\phi_\Theta]- \frac{m}{2} + \frac{1}{2}(\mu_\Theta- \mu_\Theta^\phi)^T \Sigma_\Theta^{-1}( \mu_\Theta-\mu_\Theta^\phi)$ \\  \hline	
			 		$\mathbb{E}_Y [ KL (p_\phi(\theta) || p(\theta)) ]$ & $T_{\phi}$ & $\frac{1}{2}  \log \left[\frac{|{\Sigma}_\Theta|}{|{\Sigma}^\phi_\Theta|}\right]  + \frac{1}{2} \tr[\Sigma_\Theta^{-1} {\Sigma}^\phi_\Theta]- \frac{m}{2} + \frac{1}{2}(\mu_\Theta- \mu_\Theta^\phi)^T \Sigma_\Theta^{-1}( \mu_\Theta-\mu_\Theta^\phi)$ \\  \hline	
			 		$\mathbb{E}_Y [ KL (p_\phi(\theta|y) || p(\theta|y)) ]$ & $D_\phi$ & $-\frac{1}{2} \log \left[\frac{|Q|}{|Q_\phi|}\right]  -  \frac{1}{2} \tr({Q}^{-1} {Q}_\phi) + \frac{m}{2} - \frac{1}{2}(\mu_\Theta-\mu_\Theta^\phi)^T Q^{-1} (\mu_\Theta-\mu_\Theta^\phi)$ \\  
			 		&& $- \frac{1}{2} \tr[(R-R_\phi)^T Q^{-1} (R-R_\phi) \Sigma_Y]$ \\  \hline	
			\end{tabular}
		\end{center}
	}
\end{table}

The distribution $p(y)$ is not available, and $\mathbb{E}_Y$ and gradients have to be estimated via sampling. Thus the so-called `density view' (assuming exact evaluation of $\mathbb{E}_{\Theta|Y}^\phi$) is also presented in table ~\ref{tab:density}. For compactness of notation,  given a vector $q$ and a matrix $U$ , $\norm{q}_U \triangleq q^T  U^{-1} q$.

\begin{table}[htbp]
	{   
		\caption{Density View}
		\label{tab:density}
		\begin{center}
			\begin{tabular}{|c|c|l|} \hline
				\bf Term & \bf Notation  & \bf Expression  \\  \hline \hline
				$-  \mathbb{E}_{\Theta|Y}^\phi [\log p(y|\theta)] $ & $\ell_{rec}$   &   $\frac{n}{2}\log{2 \pi } + \frac{1}{2} \log|S|  + \frac{1}{2}   \norm{ A (R_\phi y + b_\phi) -y }_S^2  $ \\
				&& $+   \frac{1}{2}\tr[A^T S^{-1} A Q_\phi]$ \\ \hline
				$KL (p_\phi(\theta|y) || p(\theta)) $ & $\ell_{reg}$ & $ \frac{1}{2} \log \left[\frac{|{\Sigma}_\Theta|}{|{Q}_\phi|}\right]  + \frac{1}{2}\norm{R_\phi y+b_\phi- \mu_\Theta}_{\Sigma_\Theta}^2 + \frac{1}{2} \tr(\Sigma_\Theta^{-1} {Q}_\phi) - \frac{m}{2}$ \\  \hline	
				$ KL (p_\phi(\theta|y) || p(\theta|y)) $ & $d_\phi$ & $- \frac{1}{2} \log \left[\frac{|Q|}{|Q_\phi|}\right]  - \frac{1}{2}\norm{(R-R_\phi) y+  (b-b_\phi)]}_Q^2 - \frac{1}{2}  \tr({Q}^{-1} {Q}_\phi) + \frac{m}{2}$ \\  \hline	
			\end{tabular}
		\end{center}
	}
\end{table}

In practical applications,  implementations consider a simpler  version, based on one random sample from $p_\phi(y|\theta)$ instead of a full expectation. This will be discussed in Section ~\ref{sec:VEINumerical}.

\begin{figure}
	\centering
	\includegraphics[width=0.8\textwidth]{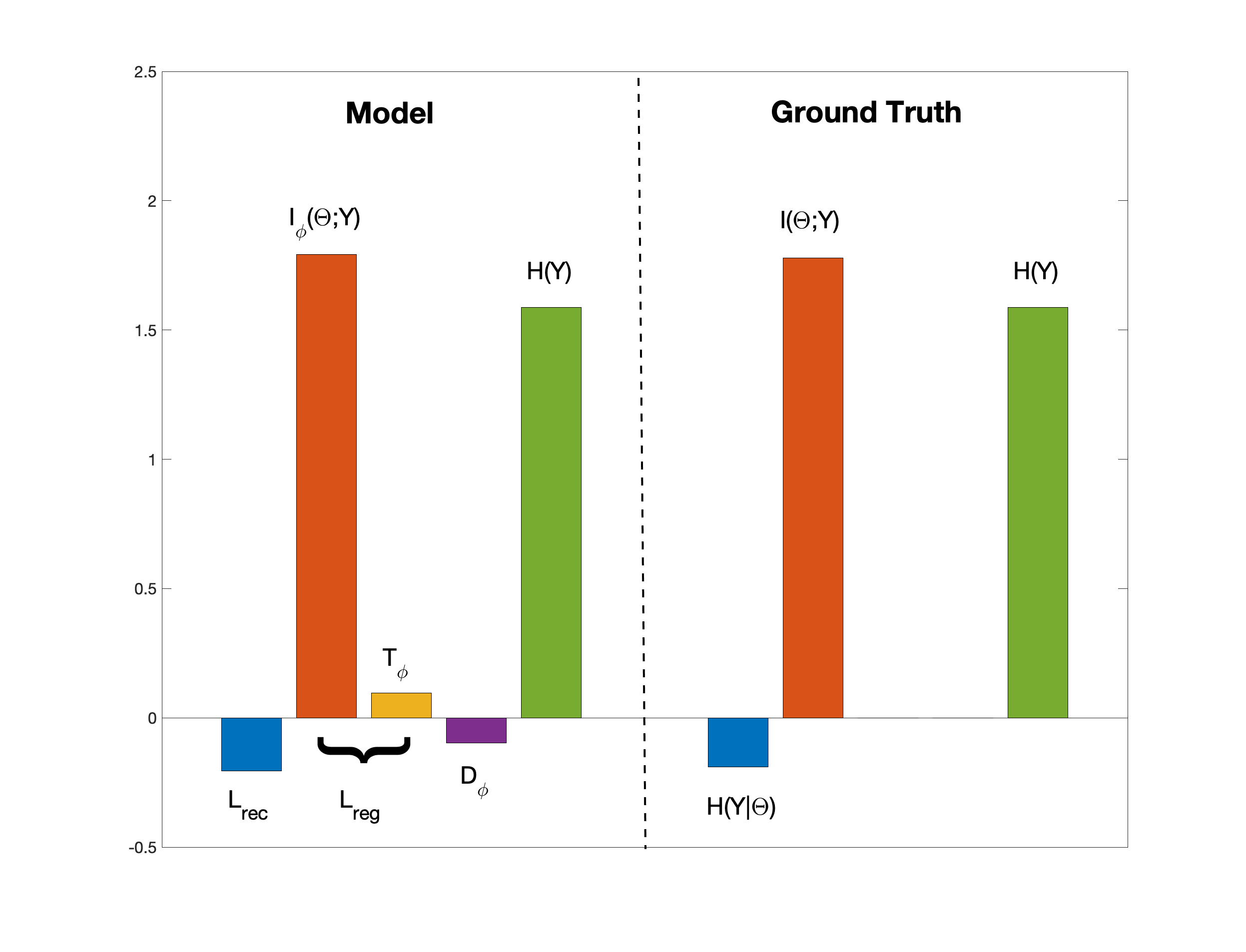}
	\caption{Budget of different terms for a sample encoding problem.}
	\label{fig:encoderbudget}
\end{figure}

Shown in Fig.~\ref{fig:encoderbudget} is the `information budget' for an encoder inferred via minimization of the ELBO (details in Section~\ref{sec:VEINumerical}).

\section{Solution of VEI Problem \& Numerical Tests}

\begin{theorem}[Solution to the Linear-Gaussian VEI Problem]\label{th:VEI}
	Given  $p(\theta)   = \mathcal{N}(\theta;\mu_\Theta,\Sigma_\Theta)$, and $p(y|\theta)  =  \mathcal{N}(y;A \theta, S)$ and defining $p_\phi(\theta|y) \triangleq \mathcal{N}(\theta;R_\phi y + b_\phi, Q_\phi)$, the solution to the VEI problem (definition~\ref{def:solution}) is 
	
	\begin{align}
	Q_\phi &=  (A^\top S^{-1}A + \Sigma_\Theta^{-1}) ^{-1} \\
	R_\phi &= Q_\phi  A^T S^{-1} \\
	b_\phi &= Q_\phi \Sigma_\Theta^{-1} \mu_\Theta .
	\end{align}

	\end{theorem}

{\bf Proof:} Appendix ~\ref{sec:VEIProof}.

\noindent It is trivially verified that these solutions correspond to the posterior computed by the Bayes rule. This should not be a surprise because the Bayes rule is a special case of the principle of minimum information~\cite{williams1980bayesian,zhu2014bayesian}.  As an  aside, Giffin \& Caticha present an interesting take on maximum entropy, minimum information and the Bayes rule~\cite{giffin2007updating}.

\subsection{Numerical investigation of VEI}\label{sec:VEINumerical}

In the numerical investigations, we use a setup that is commonly used in practical problems. Thus we do not assume that $p(y)$ is known explicitly, and instead work with samples drawn from $p(y)$. Specifically, 1024 {\em i.i.d.} samples are drawn from  $\theta \sim \mathcal{N}(0_{2 \times 1},I_{2 \times 2})$. The data samples are generated by parsing the $\theta$ samples through $y \sim \mathcal{N}(A \theta, S)$, where $A = [1 \ \  0.6]$ and $S = 0.04$.  Therefore, $\mu_Y = 0$ and $\Sigma_Y = 1.4$, though these are not provided to the code explicitly. The Adam optimizer is used in pytorch, with a learning rate of 0.001 and a batch size of 32. A total of 500 epochs were performed. 

Given $\theta \in \reals^2$ and $y \in \reals$, the parameter vector $\phi \in \reals^7$, and the unknowns are expressed as:
\begin{equation}
R_\phi = \left[
\begin{array}{c}
\phi_1 \\
 \phi_2
\end{array}
\right]    \    \     ;     \     \ b_\phi = \left[
\begin{array}{c}
\phi_3 \\
\phi_4
\end{array}
\right]  \     \ ; \      \ 
	Q_\phi = \left[
	\begin{array}{cc}
	\log(\phi_5) &    0.25 \tanh(\phi_7) \log (\phi_5 + \phi_6) \\
  0.25 \tanh(\phi_7) \log (\phi_5 + \phi_6) & \log (\phi_6)\\
	\end{array}
	\right] ,   
\end{equation}
where the parameterization of the covariance matrix is  done in a way to ensure that
\begin{equation}
	Q_\phi = \left[
\begin{array}{cc}
\sigma_1^2 &    \rho \sigma_1 \sigma_2 \\
 \rho \sigma_1 \sigma_2  & \sigma_2^2 \\
\end{array}
\right] 
\end{equation}
is symmetric positive definite.

The loss function is approximated by replacing $\mathbb{E}_Y$ by an average over the mini-batch and $\mathbb{E}_{\Theta|Y}^\phi$ - as in almost all practical implementations - by using one random sample, i.e.
$$
\mathbb{E}_{\Theta|Y}^\phi [ f(\theta) ] \approx  f(R_\phi y + b_\phi + C_\phi \epsilon),
$$
where $\epsilon \sim  N(0,I_{2 \times 2})$ and $ C_\phi C_\phi^T = Q_\phi$, with
\begin{equation}
C_\phi = \left[
\begin{array}{cc}
\sigma_1 &    0 \\
\rho \sigma_2  & \sigma_2\sqrt{1-\rho^2} \\
\end{array}
\right] 
\end{equation}

Note that this is non-standard since it is typical to use a diagonal $Q_\phi$.

Thus,
\begin{align}\label{eq:samples}
L_{rec} &\approx  L_{rec}^\ast \triangleq \frac{1}{N} \sum_{i=1}^{N} \left[  \frac{n}{2}\log{2 \pi } + \frac{1}{2} \log|S|  + \frac{1}{2}   \norm{ A (R_\phi y_i + b_\phi + C_\phi \epsilon_i) -y_i }_S^2 \right] \\
L_{reg} & \approx L_{reg}^\ast \triangleq \frac{1}{N}\sum_{i=1}^{N}  \left[  \frac{1}{2} \log \left[\frac{|{\Sigma}_\Theta|}{|{C}_\phi C_\phi^T|}\right]  + \frac{1}{2}\norm{R_\phi y_i+b_\phi + C_\phi \epsilon_i - \mu_\Theta}_{\Sigma_\Theta}^2  -  \frac{1}{2}\norm{ C_\phi \epsilon_i}_{C_\phi C_\phi^T}^2 \right].
\end{align}

\begin{figure}
	\centering
	\includegraphics[width=0.5\textwidth]{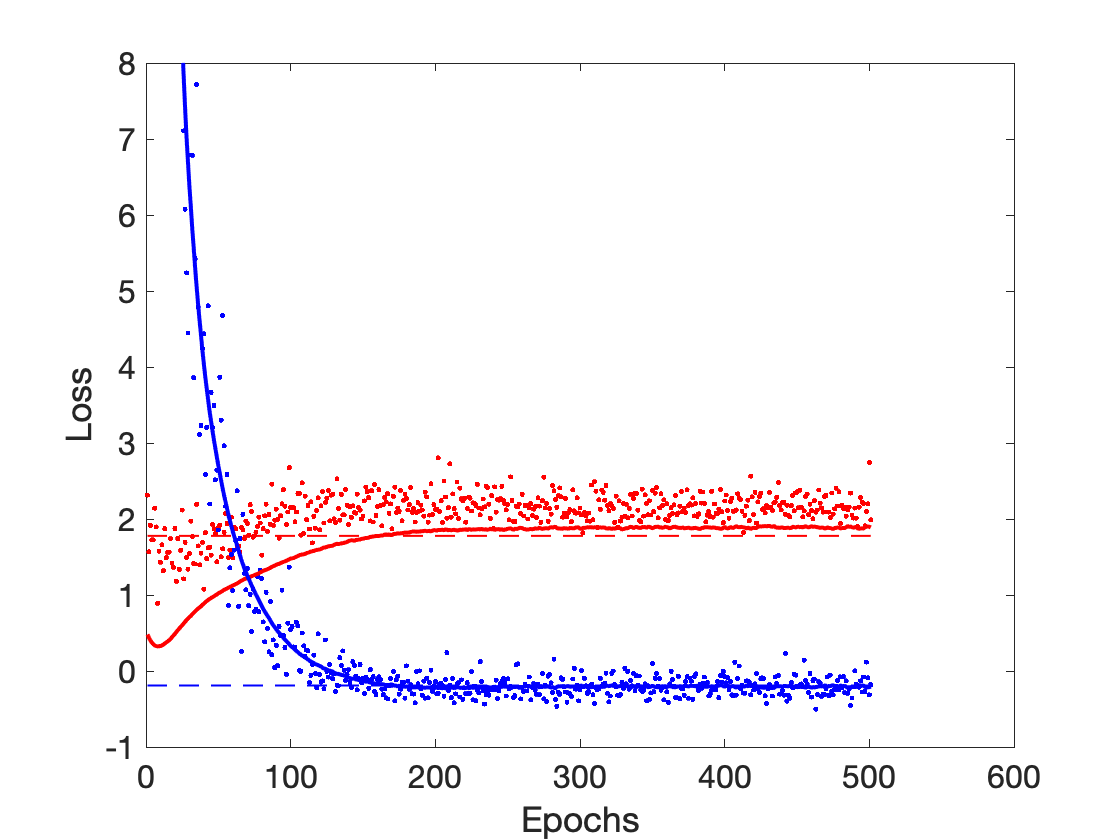}
	\caption{Convergence of  $L_{reg}^\ast$ (red dots ) and $L_{rec}^\ast$ (blue dots) during the optimization. Also shown are $L_{reg}$ (red solid line) and $L_{rec}$ (blue solid line) computed using the expressions in Table~\ref{tab:full}. Also shown are $I(Y;\Theta)$ (red dashed line) and $H(Y|\Theta)$ (blue dashed line) of the ground truth distribution.}
	\label{fig:VEI1}
\end{figure}

The convergence of the loss function components is shown in Figure~\ref{fig:VEI1}.

\begin{figure}
	\centering
	\includegraphics[width=0.5\textwidth]{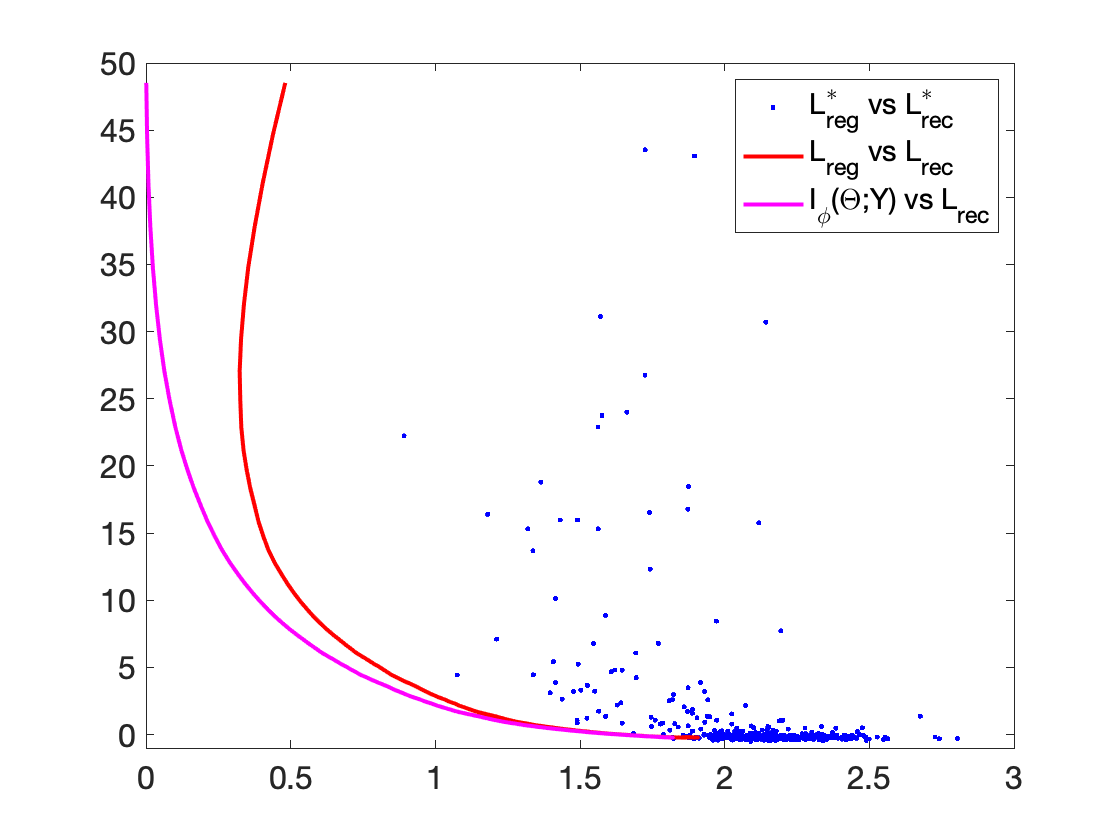}
	\caption{Evolution of different quantities during the optimization}
	\label{fig:VEI2}
\end{figure}

The difference between the sample-based loss function components (symbols) and the  analytically integrated loss function components (determined at each epoch as a post processing using the expressions in Table~\ref{tab:full}) are shown in Figure~\ref{fig:VEI2}. Also shown is $I_\phi(\Theta; Y)$.

\begin{figure}
	\centering
	\includegraphics[width=0.48\textwidth]{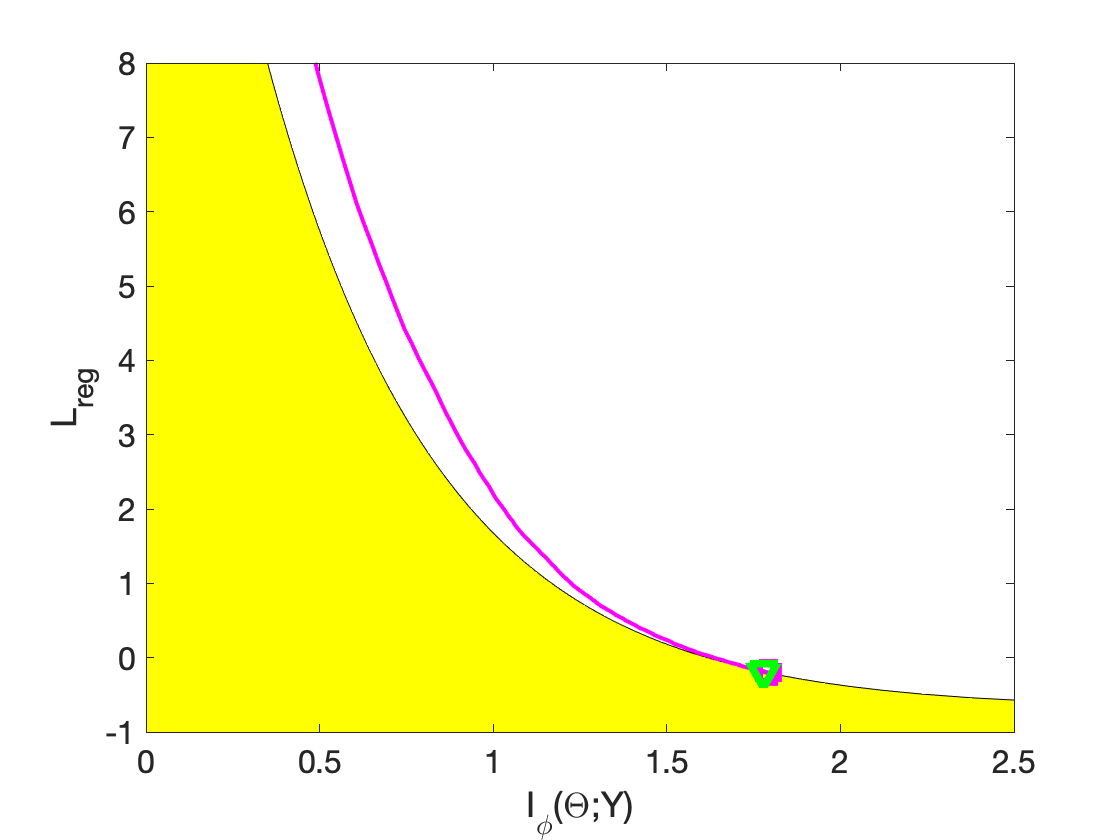}
		\includegraphics[width=0.48\textwidth]{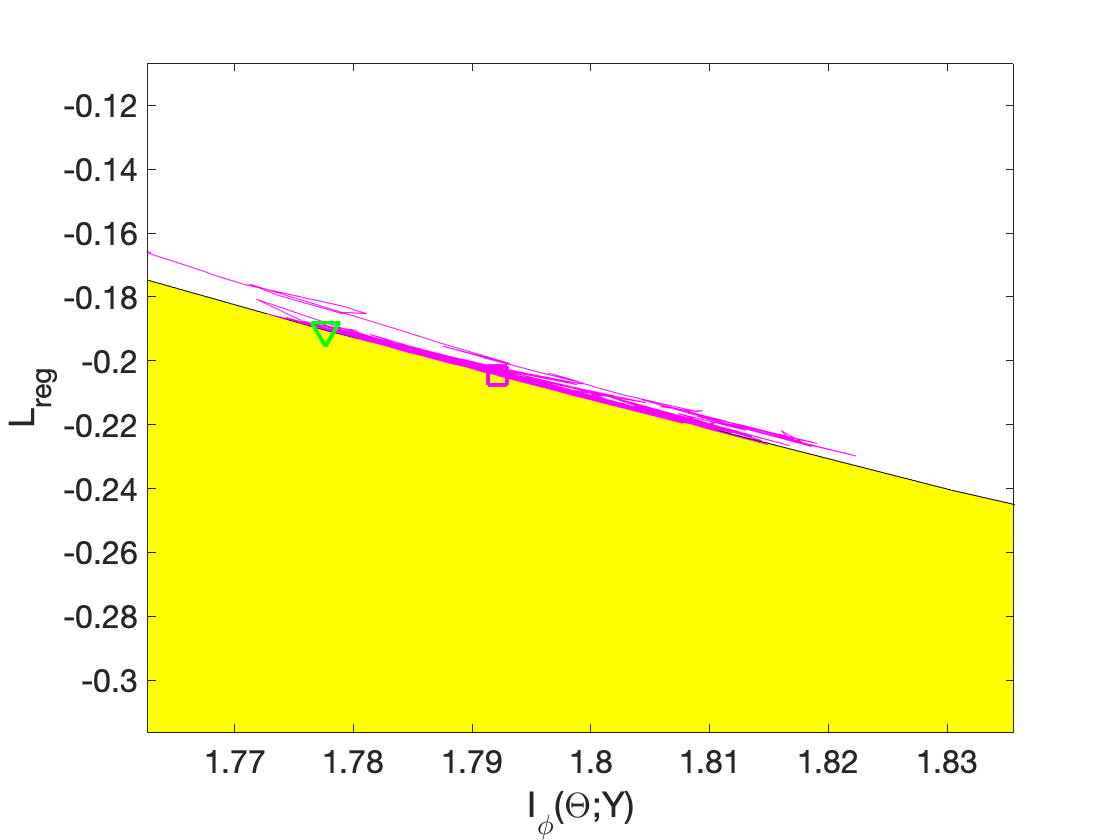}
	\caption{Evolution of $I_\phi(\Theta;Y)$ and $L_{reg}$  during the optimization (magenta line). The square symbol denotes the final solution and the green symbol denotes the ground truth. The Yellow shaded region is unachievable and is determined by the Blahut Arimoto algorithm from Rate Distortion theory.}
	\label{fig:VEI3}
\end{figure}

The evolution of $I_\phi(\Theta;Y)$ and $L_{reg}$ are shown in the RD-plane in Figure~\ref{fig:VEI3}. The RD threshold is computed using Eq.~\ref{eq:ba}.

\begin{figure}
	\centering
	\includegraphics[width=0.5\textwidth]{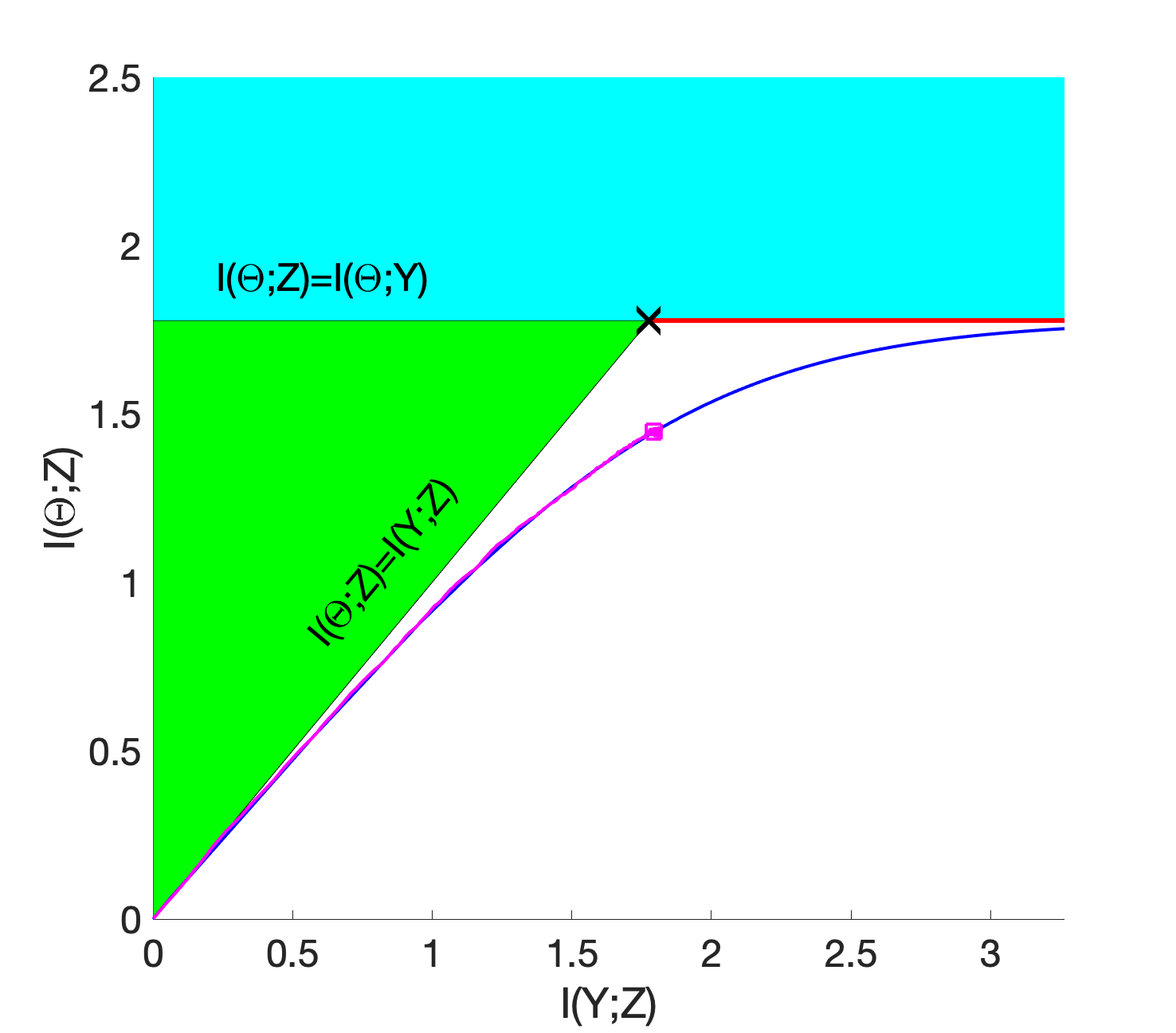}
	\caption{Evolution of the optimization iterations (magenta line)  in the information plane. The square symbol denotes the final solution  The blue line is the information bottleneck solution (for different $\beta$). The shaded regions correspond to infeasible regions defined by the data processing inequality. The red line signifies a sufficient statistic for $\Theta$ and the black cross denotes the Minimal sufficient statistic}
	\label{fig:VEI4}
\end{figure}

\subsection{Minimal Sufficient Statistic and Information Bottleneck}

To cast our problem in the classical Information Bottleneck (IB) setting (see section~\ref{subsec:IB}), we use a slight change of notation. Instead of representing the encoder by $p_\phi(\theta|y)$ and the induced joint distribution by $p_\phi(y,\theta)$, we introduce a  random variable Z and denote the encoder and the induced joint distribution by $p(z|y)$ and $p(y,z) = p(z|y) p(y)$ instead of $p_\phi(\theta|y)$ and $p_\phi(\theta,y)$, respectively. Therefore, in the context of IB, we have a linear Gaussian markov chain $\Theta \to Y \to Z$ which is generated by  $p(\theta) = \mathcal{N}(\theta;\mu_\Theta,\Sigma_\Theta)$, $p(y|\theta)  =  \mathcal{N}(y ; A \theta, S)$, and  
$p(z|y)  =  \mathcal{N}(z;R_\phi y + b_\phi, Q_\phi)$. 

Then, the joint distribution is
\begin{equation}\label{eq:3Markov}
 \left[ \begin{array}{c} \Theta \\  Y \\ Z \end{array} \right] \sim  \mathcal{N}\left(\left[ \begin{array}{c} \mu_\Theta \\ \mu_Y \\ R_\phi \mu_Y + b_\phi \end{array} \right],
\left[
\begin{array}{ccc}
\Sigma_{\Theta} &    \Sigma_\Theta A^T &  \Sigma_\Theta A^T R_\phi^T\\
A \Sigma_{\Theta} &  \Sigma_Y & \Sigma_Y R_\phi^T \\
R_\phi A \Sigma_{\Theta} & R_\phi \Sigma_Y & R_\phi \Sigma_Y R_\phi^T  + Q_\phi\\
\end{array}
\right]\right).
\end{equation}

The evolution of the optimization in the information plane is shown in Figure~\ref{fig:VEI4}. Note that $Z$ is a sufficient statistic for $\Theta$ if  $ I(\Theta ;Z) = I(\Theta ;Y)$. This is given by the red line in Figure~\ref{fig:VEI4}.

$Z$ is a minimal sufficient statistic if it satisfies
\begin{equation}\label{eq:MSS}
\min_{R_\phi,Q_\phi}   I(Y;Z)  \  \ \textrm{subject to} \ \ I(\Theta ;Z) = I(\Theta ;Y).
\end{equation}

Due to the data processing inequality, $I(Y;Z) \geq I(\Theta ;Z)$ and $I(\Theta;Y) \geq I(\Theta;Z)$ and therefore the best compression one can hope to achieve will yield $I(Y;Z) = I(\Theta;Z)$.  
Therefore, with the sufficiency constraint,  we have  

$$ I(Y;Z) =  I(Y ;\Theta ) = I(\Theta ; Z).$$

The minimal sufficient statistic is shown as a black cross in Figure~\ref{fig:VEI4}.

Adapting the information bottleneck (IB) problem~\cite{tishby2000information} as a Lagrangian version of Eq.~\ref{eq:MSS}, we have
\begin{equation}
\min_{R_\phi,Q_\phi}   I(Y;Z)  - \beta I(\Theta ;Z).
\end{equation}

The numerical solution to the above problem (using Eq.~\ref{eq:ba_IB}) is shown as the blue line in Figure~\ref{fig:VEI4}. The  numerical solution  when converges around $I(Y;Z) \approx I(Y;\Theta)$ as expected.  

It is interesting that the optimization proceeds on the IB pareto front, computed using the Blahut-Arimoto algorithm~\ref{eq:ba_IB}. While this trajectory can be explained to a certain degree as a consequence of the fact that SGD (and variants) are  endowed with variational inference properties~\cite{mandt2017stochastic},  it is remarkable that the optimization proceeds precisely on the IB line. Even more interestingly, the same type of optimality was noticed even when -2- sample points were used with a batch size of -1-.  It was confirmed that the same optimization path was followed in the information plane for different number of data points, ranging from 1024 to merely 2. The behavior was replicated when the optimizer was changed to Stochastic Gradient Descent (SGD).

\begin{figure}
	\centering
	\includegraphics[width=0.6\textwidth]{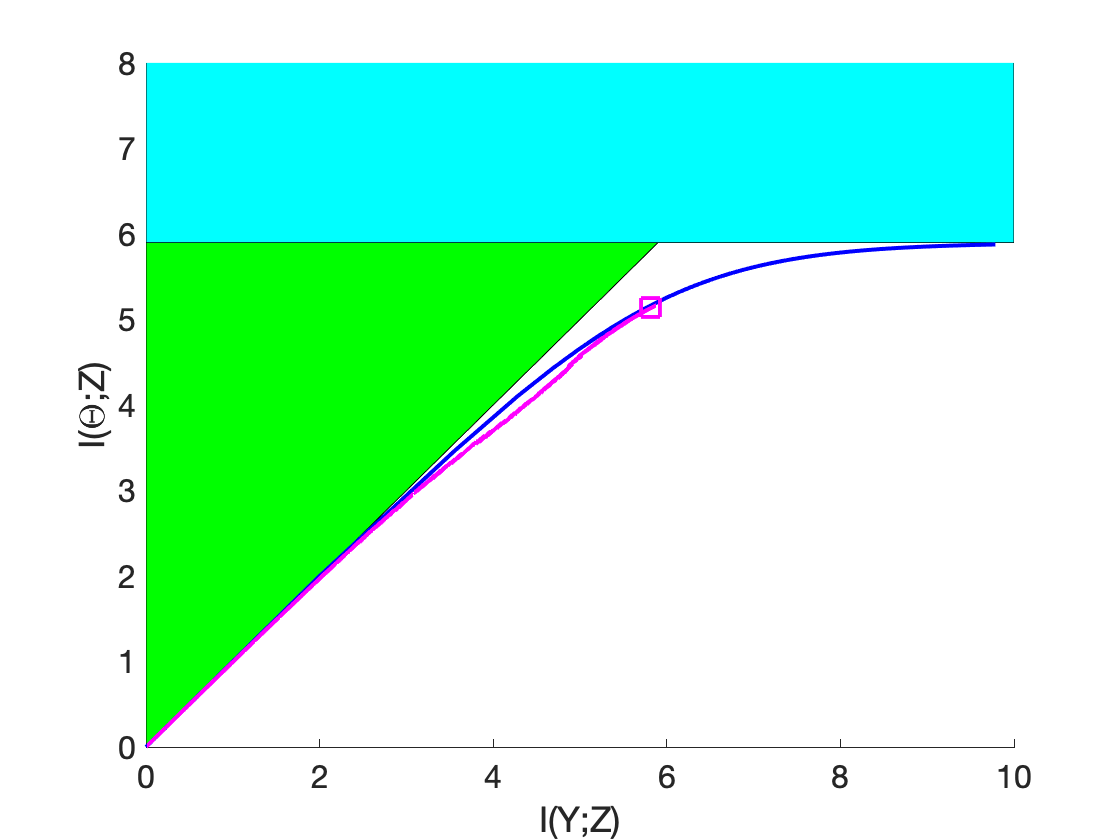}
	\caption{Evolution of the optimization iterations (magenta line)  in the information plane for the $n=4$ problem}
	\label{fig:VEI5}
\end{figure}

To see whether this behavior holds, the likelihood was changed to 
$$
A= \left[
\begin{array}{c c}
1 & 0.6 \\
3.2 & -2 \\
4 & 1 \\
3.1 & -1 \\
\end{array}
\right]
$$
(thus, $n=4$ instead of $n=1$ in the previous problem). Again, a similar behavior was noticed for the SGD trajectory as shown in Figure~\ref{fig:VEI5}.

\section{Solution of VES problems and connection to Rate Distortion theory}

\begin{theorem}[Solution to the Linear-Gaussian VES Problem]\label{th:VES}
	Given  $p(y)   =  \mathcal{N}(y; \mu_Y,\Sigma_Y)$, and $p(y|\theta)  =  \mathcal{N}(y; A \theta, S)$ and defining $p_\phi(\theta|y) \triangleq \mathcal{N}(\theta;R_\phi y + b_\phi, Q_\phi)$, the solution to the VES problem (definition~\ref{def:solution}) satisfies 
	
	\begin{align}
 A  Q_\phi A^T   &= S  - S  \Sigma_Y^{-1} S  \\
	R_\phi &= Q_\phi  A^T S^{-1} \\
 A (R_\phi \mu_Y + b_\phi)  &= A^\top S^{-1} \mu_Y.
	\end{align}
	
	It is noted that $R,b,Q$ from the generating distribution (Eq.~\ref{eq:generating}) satisfies the above equations.
	
	In particular, if $A$ has full row rank, the above equations have an explicit solution:

\begin{align}
Q_\phi &= A^+ (S  - S  \Sigma_Y^{-1} S) A^{+T}  \\
R_\phi &= Q_\phi A^T S^{-1} = A^+ (I-S \Sigma_Y^{-1})\\
b_\phi &=  A^+ S \Sigma_Y^{-1} \mu_Y.
\end{align}

The marginal distribution is given by

\begin{align}
\mu_\Theta^\phi &=  A^+ \mu_Y  \\
\Sigma_\Theta^\phi &=  A^+ \left( \Sigma_Y  - S \right) A^{+T}.
\end{align}

Also, $I_\phi(\Theta;Y) = I(\Theta;Y)$ and $H_\phi(Y|\Theta) = H(Y|\Theta)$.
	
\end{theorem}

{\bf Proof} Check section~\ref{sec:bVESProof} (for $\beta=1$).  

When $A$ has full row rank, the relationship of the optimal solution to the original generating distribution is given by
\begin{align*} \label{eq:VESSimilar}
	\Sigma_\Theta^\phi = A^+ A \Sigma_\Theta  A^T A^{+T}   \ \ ; \ \ 		 	Q_\phi = A^+ A Q  A^T A^{+T}  \\
		 \mu_\Theta^\phi =         A^+ A \mu_\Theta \ \ ; \ \  R_\phi = A^+ A R  \  \ ; \    \ b_\phi = A^+ A b. 
\end{align*}


\begin{theorem}[Solution to the Linear-Gaussian $\beta-$VES Problems]\label{th:bVES}
	Given  $p(y)   =  \mathcal{N}(y; \mu_Y,\Sigma_Y)$, and $p(y|\theta)  =  \mathcal{N}(y; A \theta, S)$ and defining $p_\phi(\theta|y) \triangleq \mathcal{N}(\theta;R_\phi y + b_\phi, Q_\phi)$, the solution to the following so-called $\beta-$VES problem (definition~\ref{def:solution}) satisfies 
	
	\begin{align}
A  Q_\phi A^T   &= \frac{S}{\beta}  - \frac{1}{\beta^2}S  \Sigma_Y^{-1} S  \\
	R_\phi &= \beta Q_\phi  A^T S^{-1} \\
	A (R_\phi \mu_Y + b_\phi)  &= A^\top S^{-1} \mu_Y.
	\end{align}

	In particular, if $A$ has full row rank, the above equations have an explicit solution:

	\begin{align}
Q_\phi &= \frac{1}{\beta}A^+ (S  - \frac{1}{\beta} S  \Sigma_Y^{-1} S) A^{+T}  \\
R_\phi &= \beta Q_\phi A^T S^{-1} = A^+ ( I -\frac{1}{\beta} S \Sigma_Y^{-1})\\
b_\phi &=   \frac{1}{\beta}  A^+ S \Sigma_Y^{-1}\mu_Y.
	\end{align}
	
	The marginal distribution is defined by
	
		\begin{align}
	\mu_\Theta^\phi &=  A^+ \mu_Y  \\
	\Sigma_\Theta^\phi &=  A^+ \left( \Sigma_Y  - \frac{S}{\beta} \right) A^{+T}.
	\end{align}
	
\end{theorem}

Proof: Check Appendix ~\ref{sec:bVESProof}.

\noindent The above  is equivalent to the solution of the following RD problem 

\begin{equation}
\min_{R_\phi,b_\phi,Q_\phi} I_\phi(\Theta; Y)  + \beta \mathbb{E}^\phi_{\Theta,Y}\left[d(y, \tilde{y}(z))\right] 
\end{equation}
with the achieved $I_\phi(\Theta;Y) = I(\Theta,Y)+\frac{n}{2} \log{\beta} $ and $\mathbb{E}^\phi_{\Theta,Y}\left[d(y, \tilde{y}(z))\right]  = H(Y|\Theta) + \frac{n}{2}\left[ \frac{1}{\beta} -1\right]$. Further $\Sigma_{Y|\Theta}^\phi =\frac{ \Sigma_{Y|\Theta} }{\beta}.$ This is shown in Appendix~\ref{sec:bVESProofRD}.

Solving for $\beta$, the rate distortion curve is obtained as
\begin{equation}
R(D) =  I(\Theta;Y) - \frac{n}{2} \log \left[1+ \frac{2}{n} (D - H(Y|\Theta)) \right].
\end{equation}

\noindent It was confirmed numerically  that the above analytical solution matches the Blahut Arimoto algorithm given by Eq.~\ref{eq:ba}. Thus, we have derived an analytical Rate distortion solution for the linear Gaussian case, with the distortion measure assumed to be the negative log likelihood.

\begin{figure}
	\centering
	\includegraphics[width=0.6\textwidth]{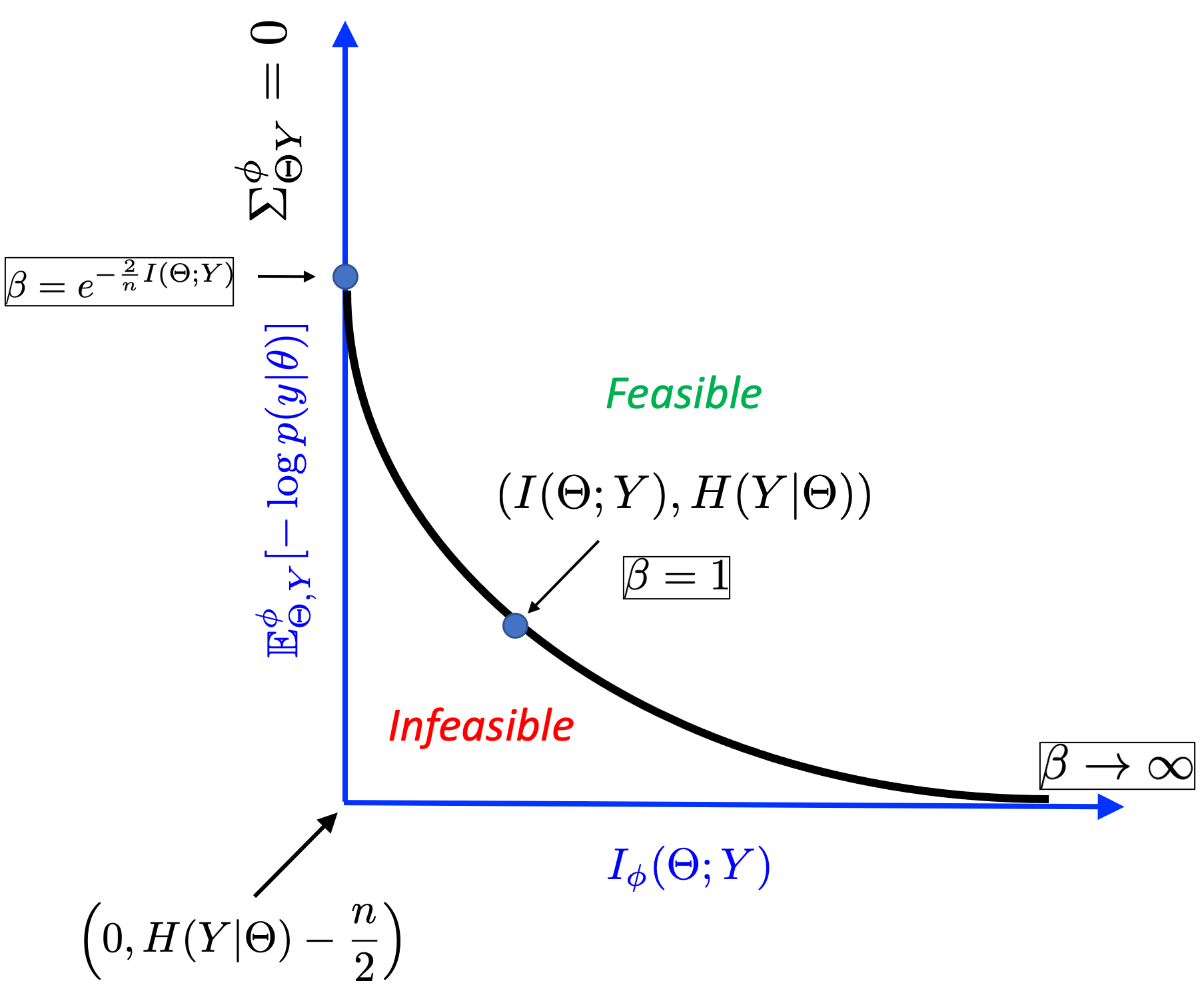}
	\caption{Schematic of the Rate distortion curve, as a solution of the $\beta-$VES problem.}
	\label{fig:RDPlane}
\end{figure}

A schematic of the Rate Distortion Pareto front is shown in Fig.~\ref{fig:RDPlane}. For the solutions on the Pareto front, the following observations can be made:

$\bullet$ When $\beta =1$, the rate and distortion match that of the generating distribution, and the encoder is similar to the generating distribution up to the matrix $A^+ A$, as shown in Eq.~\ref{eq:VESSimilar}.

$\bullet$ The minimum achievable distortion occurs when $\beta \rightarrow \infty$, a limit at which $\mathbb{E}^\phi_{\Theta,Y}\left[d(y, \tilde{y}(z))\right]  = H(Y|\Theta) - \frac{n}{2}$, and $I_\phi(\Theta;Y) = \infty$. At this limit, $R_\phi = A^+ , Q_\phi = 0, b_\phi =0$.

$\bullet$ At the other extreme, the minimum rate of 0 is achieved when $\beta = \exp \left[-\frac{2}{n} I(\Theta;Y) \right]$ and $\mathbb{E}^\phi_{\Theta,Y}\left[d(y, \tilde{y}(z))\right] = H(Y|\Theta) + \frac{n}{2} \exp \left( \left[\frac{2}{n} I(\Theta;Y) \right] -1\right)$.

Indeed, the above solutions are only valid when $Q_\phi$ and $\Sigma_\Theta^\phi$ are positive semi-definite. This requires additional conditions for validity, especially when $\beta < 1$. For instance, the relationship of $\Sigma_\Theta^\phi$ to the generating distribution $\Sigma_\Theta$ is 
$$
\Sigma_\Theta^\phi = A^+ A \Sigma_\Theta  A^T A^{+T}  - A^+ \left(  \frac{S}{\beta}-S\right) A^{+T},
$$
and thus a strong condition for positive semi-definiteness requires the analysis of the eigenvalues of the two matrices above.

\section{Variational Autoencoders}
In our Linear Gaussian variational autoencoder problems,  we are given the data distribution $p(y)   \triangleq  \mathcal{N}(y; \mu_Y,\Sigma_Y)$. The goal is to find the encoder  $p_\phi(\theta|y)  =  \mathcal{N}(\theta; R_\phi y + b_\phi , Q_\phi)$ and a decoder $p_\psi(y|\theta)  =  \mathcal{N}(y;A_\psi \theta , S_\psi)$.

The induced joint distribution of the data and  encoder  is
\begin{equation}
p_{\phi}(y,\theta)  =   \mathcal{N}\left(\left[ \begin{array}{c} \theta \\ y \end{array} \right];\left[ \begin{array}{c} R_\phi \mu_Y+ b_\phi \\ \mu_Y \end{array} \right],
\left[
\begin{array}{cc}
R_\phi \Sigma_Y R_\phi^T + Q_\phi &  R_\phi  \Sigma_Y \\
\Sigma_{Y} R_\phi^T &  \Sigma_Y
\end{array}
\right]\right).
\end{equation}

Define $\mu_\Theta^\phi \triangleq R_\phi \mu_Y + b_\phi$ and $\Sigma_\Theta^\phi \triangleq R_\phi \Sigma_Y R_\phi^T + Q_\phi$.

The induced joint distribution of the decoder is
\begin{equation}
p_\psi(\theta, y)  =   \mathcal{N}\left(\left[ \begin{array}{c} \theta \\ y \end{array} \right];\left[ \begin{array}{c} \mu_\Theta^\psi \\ A_\psi  \mu_\Theta^\psi  \end{array} \right],
\left[
\begin{array}{cc}
\Sigma_\Theta^\psi  &    \Sigma_\Theta^\psi A_\psi^T \\
A_\psi \Sigma_{\Theta}^\psi & A_\psi \Sigma_\Theta^\psi  A_\psi^T + S_\psi
\end{array}
\right]\right),
\end{equation}
where for VAEI, $\mu_\Theta^\psi = \mu_\Theta; \Sigma_\Theta^\psi = \Sigma_\Theta$, and for VAES, $\mu_\Theta^\psi = \mu_\Theta^\phi; \Sigma_\Theta^\psi = \Sigma_\Theta^\phi$ as can be seen from definition~\ref{def:solution}.

Starting with the Bayes rule,

\begin{align*}
p_\psi(y) &=  \frac{p_\psi(y|\theta) p(\theta)}{p_\psi(\theta|y)}  =   \frac{p_\psi(y|\theta) p(\theta)}{p_\psi(\theta|y)}  \frac{p_\phi(\theta|y)}{p_\phi(\theta|y)}\\
-\log p_\psi(y) &=   -\log p_\psi(y|\theta) +\log \left[ \frac{ p_\phi(\theta|y)}{p(\theta)}  \right] - \log \left[ \frac{ p_\phi(\theta|y)}{p_\psi(\theta|y)} \right] \\
\end{align*}
Let's now take an expectation over $p_\phi(\theta,y) = p(y) p_\phi(\theta|y)$
\begin{align*}
-\mathbb{E}_Y \log p_\psi(y) &=  - \mathbb{E}_Y \mathbb{E}_{\Theta|Y}^\phi [\log p_\psi(y|\theta)]+ \mathbb{E}_Y \mathbb{E}_{\Theta|Y} \log \left[ \frac{ p_\phi(\theta|y)}{p(\theta)}  \right]  -  \mathbb{E}_Y [ KL (p_\phi(\theta|y) || p_\psi(\theta|y)) ]  .\\
&= L_{rec} + L_{reg} + D_{\phi ,\psi} \\
&= \mathcal{L}(\phi,\psi) + D_{\phi ,\psi}, \\
\end{align*}
where $\mathcal{L}(\phi,\psi)$ is the (negative) ELBO.

\subsection{Variational Autoencoder Inference}
For this problem,

\begin{align*}
L_{rec}&= - \mathbb{E}_Y \mathbb{E}_{\Theta|Y} [\log p_\psi(y|\theta)]  \\
&= \frac{n}{2}\log{2 \pi} + \frac{1}{2} \log|S_\psi| + \frac{1}{2}(\mu_Y - A_\psi \mu_\Theta^\phi)^T S_\psi^{-1} (\mu_Y-A_\psi \mu_\Theta^\phi) \\
&+   \frac{1}{2}  \tr\left[S_\psi^{-1} \Sigma_Y  - 2 S_\psi^{-1} A_\psi  R_\phi \Sigma_Y  +  S_\psi^{-1} A_\psi \Sigma_\Theta^\phi A_\psi^T  \right] \\
L_{reg} & = \mathbb{E}_Y [ KL (p_\phi(\theta|y) || p(\theta)) ] \\
&= \frac{1}{2} \log \left[\frac{|{\Sigma}_\Theta|}{|{Q}_\phi|}\right]  + \frac{1}{2} \tr(\Sigma_\Theta^{-1} {\Sigma}^\phi_\Theta) - \frac{m}{2} + \frac{1}{2}[(\mu_\Theta- \mu_\Theta^\phi)^T \Sigma_\Theta^{-1}( \mu_\Theta-\mu_\Theta^\phi)]  \\
\end{align*}

\begin{theorem}[Solution to the Linear-Gaussian VAEI Problem]\label{th:VAEI}
	Given  $p(y)  =  \mathcal{N}(y;\mu_Y,\Sigma_Y), p(\theta)  =  \mathcal{N}(\theta; \mu_\Theta, \Sigma_\Theta)$ and defining $p_\phi(\theta|y) \triangleq \mathcal{N}(\theta;R_\phi y + b_\phi, Q_\phi)$, and $p_\psi(y|\theta) \triangleq \mathcal{N}(y;A_\psi \theta, S_\psi)$, the solution to the VAEI problem (definition~\ref{def:solution}) satisfies

	\begin{align*}  
	Q_\phi &=  (A_\psi ^\top S_\psi ^{-1} A_\psi  + \Sigma_\Theta^{-1}) ^{-1} \\
	R_\phi &= Q_\phi  A_\psi^T S_\psi^{-1} \\
	b_\phi &= Q_\phi \Sigma_\Theta^{-1} \mu_\Theta \\
	A_\psi &= (\mu_Y \mu_\Theta^{\phi T} +  \Sigma_Y R_\phi^T)( \mu_\Theta^\phi \mu_\Theta^{\phi T} +  \Sigma_\Theta^\phi  )^{-1}  \\
	S_\psi &=  \Sigma_Y (I - R_\phi^T A_\psi^T)   + (\mu_Y   -A_\psi \mu_\Theta^{\phi}) \mu_Y^T.
	\end{align*}
	
	\end{theorem}

Proof: Check Appendix ~\ref{sec:VAEIProof}

\begin{figure}
	\centering
	\includegraphics[width=0.6\textwidth]{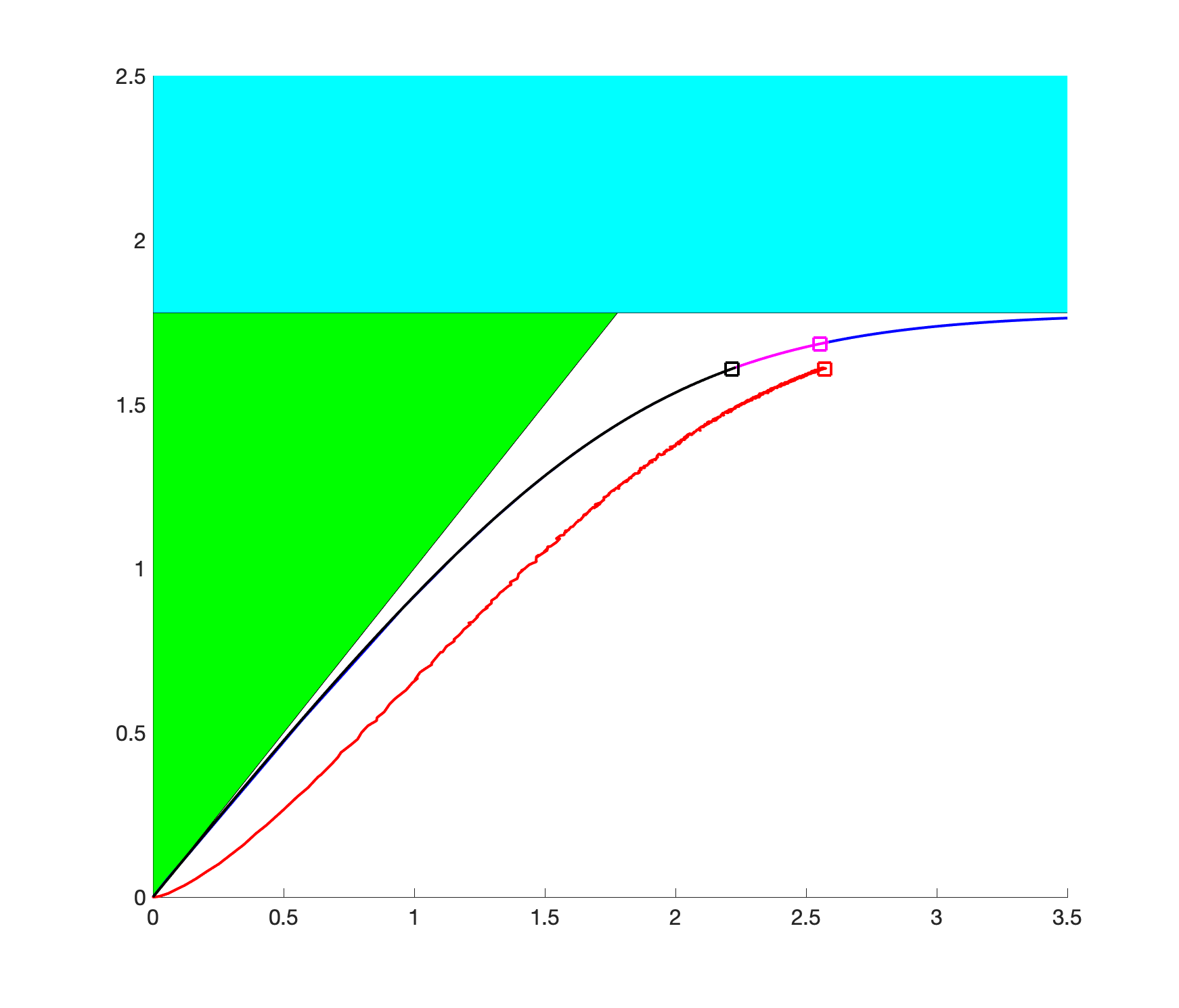}
	\caption{Evolution of the optimization iterations  in the information plane. Magenta line: $I(Y;Z) \ \ vs  \ \  I(\Theta;Z)$ ; Black line: $I(Y;\tilde{Y}) \ \ vs  \ \  I(\Theta;\tilde{Y})$ ; Red line : $I(Z;\tilde{Y}) \ \ vs  \ \  I(\Theta;\tilde{Y})$; Blue line : IB solution }
	\label{fig:VAEI1}
\end{figure}

\begin{figure}
	\centering
	\includegraphics[width=0.6\textwidth]{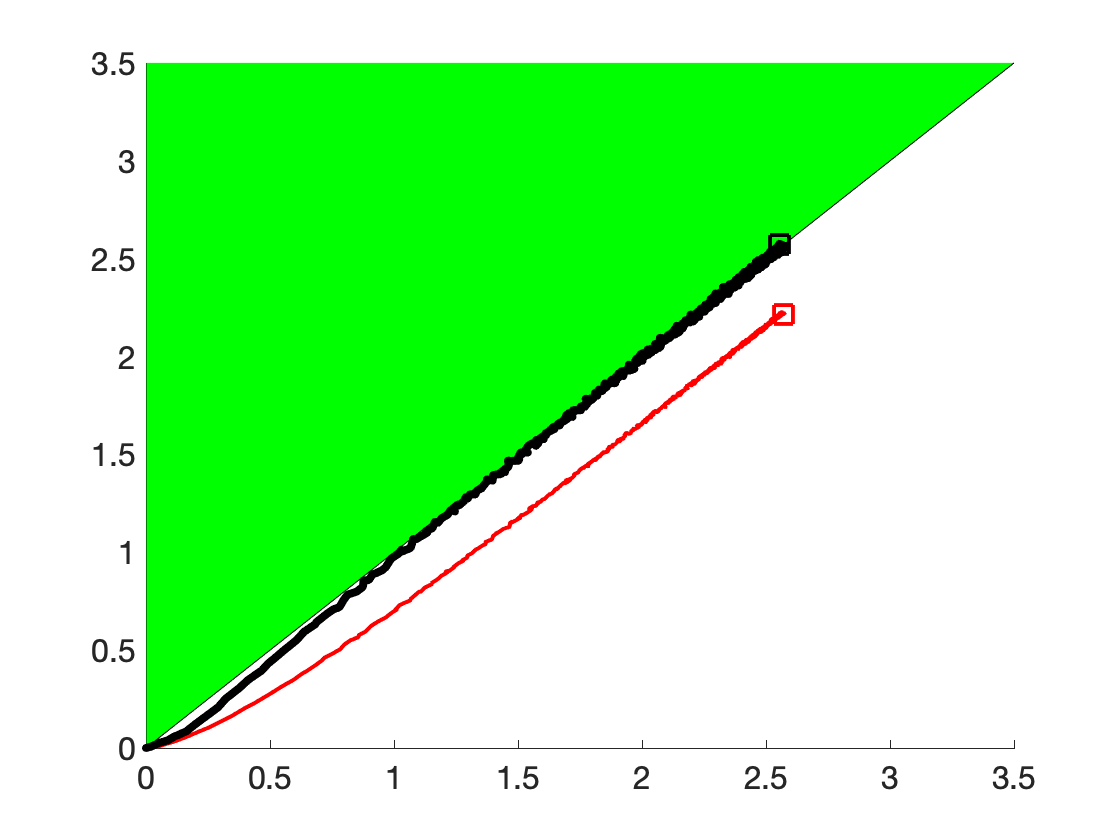}
	\caption{Evolution of the optimization iterations  in the information plane. Red line: $I(Z;\tilde{Y}) \ \ vs  \ \  I(Y;\tilde{Y})$ ; Black line: $I(Y;Z) \ \ vs  \ \  I(Z;\tilde{Y})$  }
	\label{fig:VAEI2}
\end{figure}

\subsection{Numerical Exploration of VAEI}

The same data generation as in the VEI (section ~\ref{sec:VEINumerical} is pursued), but of course, $A_\psi$ and $S_\psi$ are uknown in this case. For a batch size N, and one sample to evaluate $\mathbb{E}_{\Theta|Y}^\phi$, the loss function is estimated as:
\begin{align*}
L_{rec} &\approx  L_{rec}^\ast \triangleq \frac{1}{N} \sum_{i=1}^{N} \left[  \frac{n}{2}\log{2 \pi } + \frac{1}{2} \log|S_\psi|  + \frac{1}{2}   \norm{ A_\psi (R_\phi y_i + b_\phi + C_\phi \epsilon_i) -y_i }_{S_\psi}^2 \right] \\
L_{reg} & \approx L_{reg}^\ast \triangleq \frac{1}{N}\sum_{i=1}^{N}  \left[  \frac{1}{2} \log \left[\frac{|{\Sigma}_\Theta|}{|{Q}_\phi|}\right]  + \frac{1}{2}\norm{R_\phi y_i+b_\phi + C_\phi \epsilon_i - \mu_\Theta}_{\Sigma_\Theta}^2  -  \frac{1}{2}\norm{ C_\phi \epsilon_i}_{Q_\phi}^2 \right].
\end{align*}

For further clarity, as in Eq.~\ref{eq:3Markov}, we introduce the latent variable $Z$ in the context of the linear Gaussian Markov chain $\Theta \to Y \to Z \to \tilde{Y}$. In this case, $\Sigma_{Z|\Theta} = R_\phi S R_\phi^T + Q_\phi$ , $ \Sigma_{\tilde{Y}|Y} = A_\psi Q_\phi A_\psi^T + S_\psi$.

Figure~\ref{fig:VAEI1} shows the evolution of the optimization iteration. As in the VEI , the magenta line tracks the IB solution, but converges to a different encoder compared to VEI, as $A, S$ are not given. There are several interesting features in this optimization, which require further exploration:

$\bullet$ The final solution (magenta square) is more sufficient and less minimal compared to the VEI solution

$\bullet$ $I(Y;Z) \ \ vs  \ \  I(Z;\tilde{Y})$ also tracks the IB solution.

$\bullet$ Figure~\ref{fig:VAEI2} shows that $I(Z;\tilde{Y}) \approx I(Y;\tilde{Y})$.

\subsection{VAES}

In this case, we are long given the data distribution $p(y)$ and the dimension of the latent variable (set to be the same as that of $\Theta$). Then,

\begin{align*}
L_{rec}&= - \mathbb{E}_Y \mathbb{E}_{\Theta|Y} [\log p_\psi(y|\theta)]  \\
&= \frac{n}{2}\log{2 \pi} + \frac{1}{2} \log|S_\psi| + \frac{1}{2}(\mu_Y - A_\psi \mu_\Theta^\phi)^T S_\psi^{-1} (\mu_Y-A_\psi \mu_\Theta^\phi) \\
&+   \frac{1}{2}  \tr\left[S_\psi^{-1} \Sigma_Y  - 2 S_\psi^{-1} A_\psi  R_\phi \Sigma_Y  +  S_\psi^{-1} A_\psi \Sigma_\Theta^\phi A_\psi^T  \right] \\
L_{reg} & = \mathbb{E}_Y [ KL (p_\phi(\theta|y) || p_\phi(\theta)) ] \\
&= \frac{1}{2} \log \left[\frac{|{\Sigma}_\Theta^\phi|}{|{Q}_\phi|}\right]  
\end{align*}

\begin{theorem}[Solution to the Linear-Gaussian VAES Problem]\label{th:VAES}
	Given  $p(y)   =  \mathcal{N}(y;\mu_Y,\Sigma_Y)$ and $m$, and defining $p_\phi(\theta|y) \triangleq \mathcal{N}(\theta; R_\phi y + b_\phi, Q_\phi)$, and $p_\psi(y|\theta) \triangleq \mathcal{N}(y; A_\psi \theta, S_\psi)$, the solution to the VAES problem (definition~\ref{def:solution}) satisfies

\begin{align*}
R_\phi &= Q_\phi A_\psi^T S_\psi^{-1}\\
A_\psi  Q_\phi A_\psi^T   &= S_\psi  - S_\psi  \Sigma_Y^{-1} S_\psi  \\
A_\psi &= (\mu_Y \mu_\Theta^{\phi T} +  \Sigma_Y R_\phi^T)( \mu_\Theta^\phi \mu_\Theta^{\phi T} +  \Sigma_\Theta^\phi  )^{-1}  \\
S_\psi &=  \Sigma_Y (I - R_\phi^T A_\psi^T)   + (\mu_Y   -A_\psi \mu_\Theta^{\phi}) \mu_Y^T\\
A_\psi ^\top S_\psi ^{-1}  A_\psi \mu_\Theta^\phi &= A_\psi ^\top S_\psi ^{-1} \mu_Y. \\
\end{align*}

\end{theorem}

Proof: Check Appendix~\ref{sec:VAESProof}

\section{Summary}

This work   detailed derivations of variational inference for linear encoders and autoencoders. Four problem statements were developed in the context of variational Bayes  : a)  Encoder inference (data, likelihood and prior distributions are given), b) Encoder search (data and likelihood are given); c) Autoencoder inference (data and prior are given); d) Autoencoder search (data and the dimensionality of the parameters is given). The first two problems seek encoders (the posterior) and the latter two seek autoencoders (the posterior and the likelihood). 

$\bullet$ A linear Gaussian setting was used in each of these cases, and analytical solutions were derived, with the overall goal of establishing a principled understanding of  approaches and algorithms. Complete derivations are provided for all of the analytical results.

$\bullet$ The variational encoder inference precisely gives the Bayesian posterior, consistent with the principle of minimum information (or maximum entropy).

$\bullet$ The variational encoder search is equivalent to a rate-distortion problem with the distortion measure being the negative log likelihood. The analytical solution to this problem is an original contribution to exact Rate distortion solutions.

$\bullet$ Similar to the  encoder inference and search, analytical solutions were derived for  autoencoder inference and search. 

$\bullet$ Working with samples from the data distribution,  convergence of  stochastic gradient descent was assessed, and the dynamics in the rate-distortion and information bottleneck planes was  discussed.

\section*{Acknowledgments}
The author acknowledges support from the AFOSR computational mathematics program (Program Manager: Dr. Fariba Fahroo). The author is grateful to Prof. Alex Gorodetsky (Univ. of Michigan) for help with rate distortion theory, and for his constant references to Jaynes.

\clearpage

\section{Appendix }

\subsection{Information measures for continuous distributions} \label{sec:continuous}
In the continuous case, the expressions for differential entropy and mutual information are natural extensions of the discrete case: $H(X) = E_X(i(x))$ and $I(X;Y) = E_{XY}(i(x;y)) = E_{XY}\left(log \left[\frac{p_{XY}(x,y)}{p_X(x) p_Y(y)}\right] \right)$.

In contrast to Shannon (discrete) entropy, however, differential entropy is less intuitive and certain inequalities that are true in the discrete case do not hold here. For instance, consider the following joint distribution $(X,Y) \sim \mathcal{N}(0,K)$, where 

$$
K= 
\begin{bmatrix}
\sigma^2 & \rho \sigma^2  \\
\rho \sigma^2 &  \sigma^2
\end{bmatrix} 
$$

$H(X) = H(Y) = \frac{1}{2}\log(2 \pi e \sigma^2)$

$H(X|Y) = \frac{1}{2}\log(2 \pi e) + \frac{1}{2}log \Sigma_{X|Y} = \frac{1}{2}\log(2 \pi e \sigma^2(1-\rho^2))$

$H(X,Y) = \log(2 \pi e \sigma^2\sqrt{1-\rho^2})$

$I(X;Y) = \frac{1}{2} \log \left(\frac{1}{1-\rho^2}\right) $

when $\sigma \rightarrow 0$, we can see that $H(X) < 0$. When $\rho \rightarrow 1$, we see that $H(X|Y) \rightarrow -\infty$ and $I(X;Y) \rightarrow \infty$.

\subsubsection{Gaussian case}
For a multi-variate normal distribution $X \sim \mathcal{N}(\mu,\Sigma)$, where $ X : \Omega \rightarrow \reals^n$, 
$$H(X) = \frac{n}{2}log(2 \pi e)+\frac{1}{2} log |\Sigma|.$$ 

Given $X  \sim   \mathcal{N}(\mu_x,\Sigma_x) $ and $Y  \sim   \mathcal{N}(\mu_y,\Sigma_y)$ wihere $ X : \Omega \rightarrow \reals^n$  and $Y : \Omega \rightarrow \reals^n$, the
cross-entropy and KL divergence are given by:
$$H(P_X,P_Y) = -E_X [\log P_Y] = \frac{n}{2} \log(2 \pi) + \frac{1}{2} \log |{\Sigma}_y| + \frac{1}{2}({\mu}_x-{\mu}_y)^T {\Sigma}_y^{-1} ({\mu}_x-{\mu}_y)+ \frac{1}{2} \Tr({\Sigma}_y^{-1} {\Sigma}_x).$$

$${KL}(P_X || P_Y) = H(P_X,P_Y) - H(X) =  \frac{1}{2} \log \left[\frac{|{\Sigma}_y|}{|{\Sigma}_x|}\right] + \frac{1}{2}({\mu}_x-{\mu}_y)^T {\Sigma}_y^{-1} ({\mu}_x-{\mu}_y)+ \frac{1}{2} \Tr({\Sigma}_y^{-1} {\Sigma}_x) - \frac{n}{2}.$$ 

Given  $X \sim   \mathcal{N}({\mu}_x,{\Sigma}_x) $ and $Y \sim   \mathcal{N}({\mu}_y,{\Sigma}_y)$ with $X : \Omega \rightarrow \reals^n$ and $Y : \Omega \rightarrow \reals^m$, 

$$I(X;Y) = \frac{1}{2} \log  \left[ \frac{|\Sigma_X|}{|\Sigma_{X|Y}|}\right] =  \frac{1}{2} \log \left[ \frac{|\Sigma_Y|}{|\Sigma_{Y|X}|}\right]$$

$$H(X|Y) = H(X) - I(X;Y) =   \frac{n}{2}log(2 \pi e)   + \frac{1}{2} \log  \left[|\Sigma_{X|Y}|\right].  $$

\subsection{Derivation of Linear Gaussian Encoder Loss}\label{sec:TableProof}

We will use the following identity
\begin{align}
\mathbb{E}_X[( A x+  b)^T {\Sigma}^{-1} (A x+ b)]=  (A \mu_X + b)^T {\Sigma}^{-1} (A \mu_X  + b) + \tr[A^T {\Sigma}^{-1} A \Sigma_X]
\end{align}

Proof:
Expanding the LHS,
\begin{align*}
& = b^T {\Sigma}^{-1} b +  \mathbb{E}_X[x^T  A^T {\Sigma}^{-1} A x]  + 2 \mathbb{E}_X[b^T {\Sigma}^{-1}A x ]\\
& = b^T {\Sigma}^{-1} b +  \mu_X^T  A^T {\Sigma}^{-1} A \mu_X  + 2 b^T {\Sigma}^{-1}A  \mu_X + \tr[A^T {\Sigma}^{-1} A \Sigma_X]\\
& = b^T {\Sigma}^{-1} b +  \mu_X^T  A^T {\Sigma}^{-1} A \mu_X  +  b^T {\Sigma}^{-1}A  \mu_X + \mu_X^T A^T  {\Sigma}^{-1} b + \tr[A^T {\Sigma}^{-1} A \Sigma_X]\\
&= b^T {\Sigma}^{-1} (A  \mu_X +b) +  \mu_X^T  A^T {\Sigma}^{-1} (A \mu_X  + b) + \tr[A^T {\Sigma}^{-1} A \Sigma_X]\\
&=  (A \mu_X + b)^T {\Sigma}^{-1} (A \mu_X  + b) + \tr[A^T {\Sigma}^{-1} A \Sigma_X].
\end{align*}

Given the above identity, we can derive explicit expression for the different terms in Table~\ref{tab:full} as follows:

$\boldsymbol{L_{rec}}$

 The reconstruction term is $L_{rec} = - \mathbb{E}_Y \mathbb{E}_{\Theta|Y} [\log p(y|\theta)] $ is

\begin{align*}
&= \frac{n}{2}\log{2 \pi } + \frac{1}{2} \log|S|  + \frac{1}{2}  \mathbb{E}_Y [  (  y - A (R_\phi y + b_\phi) )^T S^{-1} (y - A  (R_\phi y + b_\phi))] + \frac{1}{2}\tr[A^T S^{-1} A Q_\phi] \\
&+ \frac{1}{2}\tr[A^T S^{-1} A Q_\phi] + \frac{1}{2}\tr[(I-A R_\phi)^{T}S^{-1} (I-A R_\phi) \Sigma_Y]\\
&= \frac{n}{2}\log{2 \pi } + \frac{1}{2} \log|S|  + \frac{1}{2}  (\mu_Y - A \mu_\Theta^\phi )^T S^{-1} (\mu_Y - A \mu_\Theta^\phi) \\
&+ \frac{1}{2}\tr[A^T S^{-1} A Q_\phi] + \frac{1}{2}\tr[(I-A R_\phi)^{T}S^{-1} (I-A R_\phi) \Sigma_Y]\\
&= \frac{n}{2}\log{2 \pi } + \frac{1}{2} \log|S|  + \frac{1}{2}  (\mu_Y - A \mu_\Theta^\phi )^T S^{-1} (\mu_Y - A \mu_\Theta^\phi) \\
&+ \frac{1}{2}\tr[A^T S^{-1} A Q_\phi] + \frac{1}{2}\tr[S^{-1} \Sigma_Y + R_\phi^T A^T S^{-1}A R_\phi \Sigma_Y - S^{-1} A R_\phi \Sigma_Y - R_\phi^T A^T S^{-1} \Sigma_Y]\\
&=  \frac{n}{2}\log{2 \pi} + \frac{1}{2} \log|S| + \frac{1}{2}(\mu_Y - A \mu_\Theta^\phi)^T S^{-1} (\mu_Y-A \mu_\Theta^\phi) \\
&+   \frac{1}{2}  \tr\left[S^{-1} \Sigma_Y  - 2 S^{-1} A  R_\phi \Sigma_Y  +  S^{-1} A \Sigma_\Theta^\phi A^T  \right]
\end{align*}

$\boldsymbol{I_\phi}$

The second term is 
$$\mathbb{E}_{\Theta|Y} \left[\frac{p_\phi(\theta|y)}{p_\phi(\theta)} \right] = I_\phi(Y;\Theta) = \frac{1}{2} \log \left[  \frac{|\Sigma_\Theta^\phi|}{|Q_\phi|} \right] .$$

$\boldsymbol{T_\phi}$

\begin{align}
\mathbb{E}_Y \mathbb{E}_{\Theta|Y} \log \left[\frac{p_\phi(\theta)}{p(\theta)} \right] = \frac{1}{2} \log \left[\frac{|\Sigma_\Theta|}{|\Sigma_\Theta^\phi|} \right] + \frac{1}{2}\mathbb{E}_Y \mathbb{E}_{\Theta|Y} [(\theta-\mu_\Theta)^T \Sigma_\Theta^{-1} (\theta-\mu_\Theta) - (\theta-\mu_\Theta^\phi)^T \Sigma_\Theta^{\phi - 1} (\theta-\mu_\Theta^\phi)  ]
\end{align}

Consider 
$$ \mathbb{E}_Y \mathbb{E}_{\Theta|Y} [(\theta-\mu_\Theta)^T \Sigma_\Theta^{-1} (\theta-\mu_\Theta)]$$
\begin{align*}
&=  \mathbb{E}_Y  [(R_\phi y + b_\phi - \mu_\Theta)^T \Sigma_\Theta^{-1} (R_\phi y + b_\phi - \mu_\Theta)  + \tr[\Sigma_\Theta^{-1}Q_\phi] ]\\
&= (\mu_\Theta  - (R_\phi \mu_Y + b_\phi )  )^T \Sigma_\Theta^{-1}( \mu_\Theta - (R_\phi \mu_Y + b_\phi ) )   + \tr[R_\phi^T \Sigma_\Theta^{-1} R_\phi \Sigma_Y] + \tr[\Sigma_\Theta^{-1}Q_\phi] \\
&= (\mu_\Theta  -  \mu_\Theta^{\phi})^T \Sigma_\Theta^{-1}( \mu_\Theta -  \mu_\Theta^{\phi})   + \tr[ \Sigma_\Theta^{-1} R_\phi \Sigma_Y R_\phi^T]  + \tr[\Sigma_\Theta^{-1}Q_\phi] \\
&= (\mu_\Theta  -  \mu_\Theta^{\phi})^T \Sigma_\Theta^{-1}( \mu_\Theta -  \mu_\Theta^{\phi})   + \tr[ \Sigma_\Theta^{-1} \Sigma_\Theta^\phi]  
\end{align*}

Therefore 
$$
\mathbb{E}_Y \mathbb{E}_{\Theta|Y} \log \left[\frac{p_\phi(\theta)}{p(\theta)} \right] = \frac{1}{2} \log \left[\frac{|\Sigma_\Theta|}{|\Sigma_\Theta^\phi|} \right] +  \frac{1}{2}  (\mu_\Theta  -  \mu_\Theta^{\phi})^T \Sigma_\Theta^{-1}( \mu_\Theta -  \mu_\Theta^{\phi})   + \frac{1}{2} \tr[\Sigma_\Theta^{-1} \Sigma_\Theta^\phi] - \frac{m}{2} 
$$

$\boldsymbol{L_{reg}}$

The combining $I_\phi + T_\phi$, we have 
$\mathbb{E}_Y [ KL (p_\phi(\theta|y) || p(\theta)) ]$

\begin{align*} 
&= \frac{1}{2} \log \left[\frac{|{\Sigma}_\Theta|}{|{Q}_\phi|}\right]  + \frac{1}{2} \tr(\Sigma_\Theta^{-1} {Q}_\phi) - \frac{m}{2} + \frac{1}{2}\mathbb{E}_Y[(R_\phi y+b_\phi- \mu_\Theta)^T \Sigma_\Theta^{-1}( R_\phi y+b_\phi-\mu_\Theta)] \\
&= \frac{1}{2} \log \left[\frac{|{\Sigma}_\Theta|}{|{Q}_\phi|}\right]  + \frac{1}{2} \tr(\Sigma_\Theta^{-1} {\Sigma}^\phi_\Theta) - \frac{m}{2} + \frac{1}{2}[(\mu_\Theta- \mu_\Theta^\phi)^T \Sigma_\Theta^{-1}( \mu_\Theta-\mu_\Theta^\phi)]  
\end{align*}

$\boldsymbol{D_\phi}$

Consider 
$$\mathbb{E}_Y [ KL (p_\phi(\theta|y) || p(\theta|y)) ]$$

\begin{align*} 
&= \frac{1}{2} \log \left[\frac{|{Q}|}{|{Q}_\phi|}\right]  + \frac{1}{2} \tr({Q}^{-1} {Q}_\phi) - \frac{m}{2} + \frac{1}{2}\mathbb{E}_Y[((R_\phi-R)y+b_\phi-b)^T {Q}^{-1}( (R_\phi-R)y+b_\phi-b) ] \\
&= \frac{1}{2} \log \left[\frac{|Q|}{|Q_\phi|}\right]  + \frac{1}{2}  \tr({Q}^{-1} {Q}_\phi) - \frac{m}{2} + \frac{1}{2}\mathbb{E}_Y[(\Delta R y+ \Delta b)^T Q^{-1} (\Delta R y+\Delta b)]\\
&= \frac{1}{2} \log \left[\frac{|Q|}{|Q_\phi|}\right]  + \frac{1}{2} \tr({Q}^{-1} {Q}_\phi) - \frac{m}{2} + \frac{1}{2}(\Delta R \mu_Y+ \Delta b)^T Q^{-1} (\Delta R \mu_Y+\Delta b)] + \frac{1}{2} \tr[(\Delta R)^T Q^{-1} (\Delta R) \Sigma_Y]
\end{align*}
Therefore the fourth term $-\mathbb{E}_Y [ KL (p_\phi(\theta|y) || p(\theta|y)) ]$ is
$$
-\frac{1}{2} \log \left[\frac{|Q|}{|Q_\phi|}\right]  -  \frac{1}{2} \tr({Q}^{-1} {Q}_\phi) + \frac{m}{2} - \frac{1}{2}(\mu_\Theta-\mu_\Theta^\phi)^T Q^{-1} (\mu_\Theta-\mu_\Theta^\phi)] - \frac{1}{2} \tr[(\Delta R)^T Q^{-1} (\Delta R) \Sigma_Y]
$$

\subsection{Proof of VEI}\label{sec:VEIProof}

The variational encoder inference problem is the solution of 
$$
\min_{R_\phi,b_\phi,Q_\phi}  L_{rec}+L_{reg}
$$

Differentiating $L_{rec}$ and $L_{reg}$, we have

\begin{align*} \frac{\partial L_{rec}}{\partial R_\phi}  &=  -A^\top S^{-1}(\mu_Y-A\mu_\Theta^\phi)\mu_Y^T - A^T S^{-1}  \Sigma_Y  +  A^T S^{-1} A R_\phi \Sigma_Y \\
\frac{\partial L_{rec}}{\partial b_\phi}  &=  -A^\top S^{-1}(\mu_Y-A\mu_\Theta^\phi)\\
\frac{\partial L_{rec}}{\partial Q_\phi}  &= \frac{1}{2} A^\top S^{-1}A\\
\frac{\partial L_{reg}}{\partial R_\phi}  &=  -\Sigma_\Theta^{-1}(\mu_\Theta-\mu_\Theta^\phi)\mu_Y^T + \Sigma_\Theta^{-1} R_\phi \Sigma_Y   \\
\frac{\partial L_{reg}}{\partial b_\phi}  &=  -\Sigma_\Theta^{-1}(\mu_\Theta-\mu_\Theta^\phi)\\
\frac{\partial L_{reg}}{\partial Q_\phi}  &=  -\frac{1}{2} Q_\phi^{-1} + \frac{1}{2} \Sigma_\Theta^{-1}
\end{align*}

Thus, we have to satisfy
\begin{align*}  
A^\top S^{-1}A +  \Sigma_\Theta^{-1} -  Q^{-1}_\phi &=0 \\
-A^\top S^{-1}(\mu_Y-A\mu_\Theta^\phi)\mu_Y^T - A^T S^{-1}  \Sigma_Y  +  A^T S^{-1} A R_\phi \Sigma_Y  -\Sigma_\Theta^{-1}(\mu_\Theta-\mu_\Theta^\phi)\mu_Y^T + \Sigma_\Theta^{-1} R_\phi \Sigma_Y &=0 \\
-A^\top S^{-1}(\mu_Y-A\mu_\Theta^\phi) -\Sigma_\Theta^{-1}(\mu_\Theta-\mu_\Theta^\phi) &=0.
\end{align*}
The first equation gives $Q_\phi=  (A^\top S^{-1}A + \Sigma_\Theta^{-1}) ^{-1} $

Assuming the 3rd equation is satisfied, the second equation is $R_\phi = Q_\phi  A^T S^{-1} $.
The third equation is
\begin{align*}
[ A^\top S^{-1}A +  \Sigma_\Theta^{-1} ] (R_\phi \mu_Y + b_\phi) &= A^T S^{-1} \mu_Y + \Sigma_\Theta^{-1} \mu_\Theta \\
Q_\phi^{-1} R_\phi \mu_Y+ Q_\phi^{-1} b_\phi &= Q_\phi^{-1} R_\phi \mu_Y + \Sigma_\Theta^{-1} \mu_\Theta\\
b_\phi &= Q_\phi \Sigma_\Theta^{-1} \mu_\Theta .
\end{align*}


\subsection{Proof of $\beta$-VES}\label{sec:bVESProof}
In this case, 

$$L_{reg} = I_\phi(Y;\Theta) = \frac{1}{2} \log \left[\frac{|R_\phi \Sigma_Y R_\phi^T + Q_\phi|}{|Q_\phi|} \right]$$

\begin{align*} \frac{\partial L_{rec}}{\partial R_\phi}  &=  -A^\top S^{-1}(\mu_Y-A\mu_\Theta^\phi)\mu_Y^T - A^T S^{-1}  \Sigma_Y  +  A^T S^{-1} A R_\phi \Sigma_Y \\
\frac{\partial L_{rec}}{\partial b_\phi}  &=  -A^\top S^{-1}(\mu_Y-A\mu_\Theta^\phi)\\
\frac{\partial L_{rec}}{\partial Q_\phi}  &= \frac{1}{2} A^\top S^{-1}A\\
\frac{\partial L_{reg}}{\partial R_\phi}  &=   (R_\phi \Sigma_Y R_\phi^T + Q_\phi)^{-1} R_\phi \Sigma_Y  \\
\frac{\partial L_{reg}}{\partial b_\phi}  &= 0 \\
\frac{\partial L_{reg}}{\partial Q_\phi}  &=  -\frac{1}{2} Q_\phi^{-1} + \frac{1}{2} (R_\phi \Sigma_Y R_\phi^T + Q_\phi)^{-1}
\end{align*}

Therefore, the equations to be solved are
\begin{align*} 
A^\top S^{-1}(\mu_Y-A\mu_\Theta^\phi) &=0 \\ 
-\beta A^\top S^{-1}(\mu_Y-A\mu_\Theta^\phi)\mu_Y^T - \beta A^T S^{-1}  \Sigma_Y  +  \beta A^T S^{-1} A R_\phi \Sigma_Y + (R_\phi \Sigma_Y R_\phi^T + Q_\phi)^{-1} R_\phi \Sigma_Y &=0\\
\beta \frac{1}{2} A^\top S^{-1}A -\frac{1}{2} Q_\phi^{-1} + \frac{1}{2} (R_\phi \Sigma_Y R_\phi^T + Q_\phi)^{-1} &=0
\end{align*}

Assuming  the first equation is satisfied, the equation for $R_\phi$ is 
$$
[\beta A^T S^{-1} A + (R_\phi \Sigma_Y R_\phi^T + Q_\phi)^{-1} ] R_\phi = \beta A^T S^{-1}
$$

The equation for $Q_\phi$ is
$$\beta A^T S^{-1} A + (R_\phi \Sigma_Y R_\phi^T + Q_\phi)^{-1} = Q_\phi^{-1}$$

Therefore,
$$R_\phi = \beta Q_\phi A^T S^{-1}$$

Let's rewrite the equation for $Q_\phi$ using the Woodbury identity

$$\beta A^T S^{-1} A + [Q_\phi^{-1}-Q_\phi^{-1}R_\phi(\Sigma_Y^{-1}+R_\phi^T Q_\phi^{-1}R_\phi)^{-1}R_\phi^T Q_\phi^{-1}] = Q_\phi^{-1}$$

Then

\begin{align*}
\beta A^T S^{-1} A -Q_\phi^{-1}R_\phi(\Sigma_Y^{-1}+R_\phi^T Q_\phi^{-1}R_\phi)^{-1}R_\phi^T Q_\phi^{-1} &= 0 \\
Q_\phi^{-1}R_\phi(\Sigma_Y^{-1}+R_\phi^T Q_\phi^{-1}R_\phi)^{-1}R_\phi^T Q_\phi^{-1} &= \beta A^T S^{-1} A \\
\beta A^T S^{-1}(\Sigma_Y^{-1}+R_\phi^T Q_\phi^{-1}R_\phi)^{-1}S^{-1} A \beta  &= \beta A^T S^{-1} A \\
\end{align*}
We can  satisfy the above equations using
\begin{align*}
S^{-1}(\Sigma_Y^{-1}+R_\phi^T Q_\phi^{-1}R_\phi)^{-1} &= \frac{I}{\beta}\\
(\Sigma_Y^{-1}+R_\phi^T Q_\phi^{-1}R_\phi) S &= \beta I \\
R_\phi^T Q_\phi^{-1}R_\phi  &= \beta S^{-1}  - \Sigma_Y^{-1} \\
\beta^2 S^{-1} A  Q_\phi A^T S^{-1}  &= \beta S^{-1}  - \Sigma_Y^{-1} \\
A  Q_\phi A^T   &= \frac{S}{\beta}  - \frac{1}{\beta^2}S  \Sigma_Y^{-1} S  \\
\end{align*}

From the above, we derive a useful relationship $$A R_\phi = \beta A Q_\phi A^T S^{-1} =  I - \frac{1}{\beta} S \Sigma_Y^{-1}$$. Therefore $$ I - A R_\phi = \frac{1}{\beta}S \Sigma_Y^{-1} \   \ ; \  \   A R_\phi \Sigma_Y  =  \Sigma_Y-\frac{S}{\beta}.   $$

This equation has multiple solutions if $A$ has full row rank (FRR), but can only be solved in a least squares sense if $A$ has full column rank. 

Let's consider the FRR case first:

In this case, the solution is 
\begin{align*}
Q_\phi &= \frac{1}{\beta}A^+ (S  - \frac{1}{\beta} S  \Sigma_Y^{-1} S) A^{+T}  \\
R_\phi &= \beta Q_\phi^\ast A^T S^{-1} = A^+ ( I -\frac{1}{\beta} S \Sigma_Y^{-1})\\
b_\phi &=   \frac{1}{\beta}  A^+ S \Sigma_Y^{-1}\mu_Y,
\end{align*}
where the last equation is a consequence of the fact that $A b_\phi =  \mu_Y - A R_\phi \mu_Y = \mu_Y - ( I-\frac{1}{\beta}S \Sigma_Y^{-1})  \mu_Y$.

\subsubsection{$L_{rec}$ and $L_{reg}$ at the Optimal solution}\label{sec:bVESProofRD}

Consider the trace term in $L_{rec}$

$$S^{-1} \Sigma_Y  - 2 S^{-1} A  R_\phi \Sigma_Y  +  S^{-1} A \Sigma_\Theta^\phi A^T$$

\begin{align*}
&= S^{-1} \Sigma_Y  - 2 S^{-1} (\Sigma_Y-\frac{1}{\beta}S)  +  S^{-1} A (R_\phi \Sigma_Y R_\phi^T + Q_\phi) A^T\\
&= S^{-1} \Sigma_Y  - 2 S^{-1} (\Sigma_Y-\frac{1}{\beta}S)  +  S^{-1} A R_\phi \Sigma_Y R_\phi^T A^T  +S^{-1} A  Q_\phi A^T\\
&= S^{-1} \Sigma_Y  - 2 S^{-1} (\Sigma_Y-\frac{1}{\beta}S)  +  (S^{-1} - \frac{1}{\beta}  \Sigma_Y^{-1})(\Sigma_Y-\frac{S}{\beta}  )+ \frac{I}{\beta}  - \frac{1}{\beta^2}  \Sigma_Y^{-1} S \\
&= S^{-1} \Sigma_Y  - 2 S^{-1} \Sigma_Y + 2\frac{I}{\beta} +  S^{-1}  \Sigma_Y - \frac{I}{\beta}  -  \frac{I}{\beta}  + \Sigma_Y^{-1} S \frac{1}{\beta^2}+ \frac{I}{\beta}  - \frac{1}{\beta^2}  \Sigma_Y^{-1} S \\
&= \frac{I}{\beta}.
\end{align*}

The expression for the induced likelihood is,
\begin{align*}
\hat{L}_{rec}  &= \frac{n}{2}\log{2 \pi} + \frac{1}{2} \log|S|  + \frac{n}{2} + \frac{n}{2}\left[\frac{1}{\beta} - 1 \right].
\end{align*}

Consider 
\begin{align*}
\Sigma_{Y|\Theta}^\phi&=\Sigma_Y-\Sigma_Y R_\phi^T \Sigma_\Theta^{\phi -1}R_\phi \Sigma_Y \\
& = \Sigma_Y-\Sigma_Y R_\phi^T  [Q_\phi^{-1}-Q_\phi^{-1}R_\phi(\Sigma_Y^{-1}+R_\phi^T Q_\phi^{-1}R_\phi)^{-1}R_\phi^T Q_\phi^{-1}] R_\phi \Sigma_Y \\
&= \Sigma_Y-\Sigma_Y R_\phi^T  [Q_\phi^{-1}-\beta^2 A^T S^{-1}(\Sigma_Y^{-1}+ \beta R_\phi^T A^T S^{-1})^{-1} S^{-1} A] R_\phi \Sigma_Y \\
&= \Sigma_Y- \beta \Sigma_Y S^{-1} A R_\phi \Sigma_Y +\beta (\Sigma_Y- S/\beta)S^{-1} (\Sigma_Y-S/\beta) \\
&= \Sigma_Y-  \beta \Sigma_Y S^{-1} ( \Sigma_Y -S/\beta) + \beta (\Sigma_Y- S/\beta)S^{-1} (\Sigma_Y-S/\beta) \\
&=  \frac{S}{\beta}
\end{align*}

Therefore,
\begin{align}
\hat{L}_{reg}   &= \frac{1}{2}\log  \left[\frac{|\Sigma_Y| }{|\frac{S}{\beta}| } \right]  =   \frac{1}{2}\log  \left[\frac{|\Sigma_Y| }{|S|} \right]+\frac{n}{2} \log \beta.
\end{align}

\subsection{Proof of VAEI}\label{sec:VAEIProof}

Here, we have

\begin{align*}
L_{rec} &= - \mathbb{E}_Y \mathbb{E}_{\Theta|Y}^\phi [\log p_\psi(y|\theta)] \\
&=  \frac{n}{2}\log{2 \pi} + \frac{1}{2} \log|S_\psi| + \frac{1}{2}(\mu_Y - A_\psi \mu_\Theta^\phi)^T S_\psi^{-1} (\mu_Y-A_\psi \mu_\Theta^\phi) \\
&+   \frac{1}{2}  \tr\left[S_\psi^{-1} \Sigma_Y  - 2 S_\psi^{-1} A_\psi  R_\phi \Sigma_Y  +  S_\psi^{-1} A_\psi \Sigma_\Theta^\phi A_\psi^T  \right] \\
L_{reg}  &= \mathbb{E}_Y [ KL (p_\phi(\theta|y) || p(\theta)) ] \\
&= \frac{1}{2} \log \left[\frac{|{\Sigma}_\Theta|}{|{Q}_\phi|}\right]  + \frac{1}{2} \tr(\Sigma_\Theta^{-1} {\Sigma}^\phi_\Theta) - \frac{m}{2} + \frac{1}{2}[(\mu_\Theta- \mu_\Theta^\phi)^T \Sigma_\Theta^{-1}( \mu_\Theta-\mu_\Theta^\phi)]  
\end{align*}

\begin{align*} \frac{\partial L_{rec}}{\partial R_\phi}  &=  -A_\psi ^\top S_\psi ^{-1}(\mu_Y-A_\psi \mu_\Theta^\phi)\mu_Y^T - A_\psi ^T S_\psi ^{-1}  \Sigma_Y  +  A_\psi ^T S_\psi ^{-1} A_\psi  R_\phi \Sigma_Y \\
\frac{\partial L_{rec}}{\partial b_\phi}  &=  -A_\psi ^\top S_\psi ^{-1}(\mu_Y-A_\psi \mu_\Theta^\phi)\\
\frac{\partial L_{rec}}{\partial Q_\phi}  &= \frac{1}{2} A_\psi ^\top S_\psi ^{-1}A_\psi\\
\frac{\partial L_{rec}}{\partial A_\psi}  &=  -S_\psi ^{-1}(\mu_Y-A_\psi \mu_\Theta^\phi) \mu_\Theta^{\phi T} - S_\psi ^{-1}  \Sigma_Y R_\phi^T +S_\psi ^{-1} A_\psi  \Sigma_\Theta^\phi \\
\frac{\partial L_{rec}}{\partial S_\psi}  &=  -\frac{1}{2}S_\psi ^{-1}(\mu_Y-A_\psi \mu_\Theta^\phi) (\mu_Y-A_\psi \mu_\Theta^\phi)^T S_\psi^{-1}  + S_\psi ^{-1}  \Sigma_Y R_\phi^T A_\psi^T  S_\psi ^{-1} - \frac{1}{2}  S_\psi ^{-1} A_\psi  \Sigma_\Theta^\phi A_\psi^T S_\psi ^{-1} -\frac{1}{2} S_\psi^{-1}  \Sigma_Y S_\psi^{-1} + \frac{1}{2} S_\psi^{-1}\\
\frac{\partial L_{reg}}{\partial R_\phi}  &=  -\Sigma_\Theta^{-1}(\mu_\Theta-\mu_\Theta^\phi)\mu_Y^T + \Sigma_\Theta^{-1} R_\phi \Sigma_Y   \\
\frac{\partial L_{reg}}{\partial b_\phi}  &=  -\Sigma_\Theta^{-1}(\mu_\Theta-\mu_\Theta^\phi)\\
\frac{\partial L_{reg}}{\partial Q_\phi}  &=  -\frac{1}{2} Q_\phi^{-1} + \frac{1}{2} \Sigma_\Theta^{-1}\\
\frac{\partial L_{reg}}{\partial A_\psi}  &=  0\\
\frac{\partial L_{reg}}{\partial S_\psi}  &=  0
\end{align*}
The equations to be satisfied are:
\begin{align*}
A_\psi ^\top S_\psi ^{-1}A_\psi  +  \Sigma_\Theta^{-1} &=  Q_\phi^{-1}\\
- A_\psi ^T S_\psi ^{-1}  \Sigma_Y  +  A_\psi ^T S_\psi ^{-1} A_\psi  R_\phi \Sigma_Y  + \Sigma_\Theta^{-1} R_\phi \Sigma_Y &=  [A_\psi ^\top S_\psi ^{-1}(\mu_Y-A_\psi \mu_\Theta^\phi) +\Sigma_\Theta^{-1}(\mu_\Theta-\mu_\Theta^\phi)]\mu_Y^T \\
A_\psi ^\top S_\psi ^{-1}(\mu_Y-A_\psi \mu_\Theta^\phi) +\Sigma_\Theta^{-1}(\mu_\Theta-\mu_\Theta^\phi)&=0\\
-S_\psi ^{-1}(\mu_Y-A_\psi \mu_\Theta^\phi) \mu_\Theta^{\phi T} - S_\psi ^{-1}  \Sigma_Y R_\phi^T +S_\psi ^{-1} A_\psi  \Sigma_\Theta^\phi  &=0 \\
S_\psi ^{-1}  \Sigma_Y R_\phi^T A_\psi^T  S_\psi ^{-1} - \frac{1}{2}  S_\psi ^{-1} A_\psi  \Sigma_\Theta^\phi A_\psi^T S_\psi ^{-1} -\frac{1}{2} S_\psi^{-1}  \Sigma_Y S_\psi^{-1}&=\frac{1}{2}S_\psi ^{-1}(\mu_Y-A_\psi \mu_\Theta^\phi) (\mu_Y-A_\psi \mu_\Theta^\phi)^T S_\psi^{-1} - \frac{1}{2}S_\psi^{-1}
\end{align*}

The first equation gives 
\begin{align*}  
Q_\phi &=  (A_\psi ^\top S_\psi ^{-1} A_\psi  + \Sigma_\Theta^{-1}) ^{-1} \\
\end{align*}

Using the 2nd and 3rd equations, we have
\begin{align*}  
R_\phi &= Q_\phi  A_\psi^T S_\psi^{-1} \\
\end{align*}  
The third equation is
\begin{align*}
[ A_\psi ^\top S^{-1}A_\psi  +  \Sigma_\Theta^{-1} ] (R_\phi \mu_Y + b_\phi) &= A_\psi ^T S_\psi ^{-1} \mu_Y + \Sigma_\Theta^{-1} \mu_\Theta \\
Q_\phi^{-1} R_\phi \mu_Y+ Q_\phi^{-1} b_\phi &= Q_\phi^{-1} R_\phi \mu_Y + \Sigma_\Theta^{-1} \mu_\Theta\\
b_\phi &= Q_\phi \Sigma_\Theta^{-1} \mu_\Theta .
\end{align*}

Rewriting equation 4, we have,
\begin{align*}
A_\psi ( \mu_\Theta^\phi \mu_\Theta^{\phi T} +  \Sigma_\Theta^\phi ) &= \mu_Y \mu_\Theta^{\phi T} +  \Sigma_Y R_\phi^T  \\
A_\psi &= (\mu_Y \mu_\Theta^{\phi T} +  \Sigma_Y R_\phi^T)( \mu_\Theta^\phi \mu_\Theta^{\phi T} +  \Sigma_\Theta^\phi  )^{-1}  \\
\end{align*}

Rewriting the last equation,
\begin{align*}
S_\psi &= -2  \Sigma_Y R_\phi^T A_\psi^T   + A_\psi  \Sigma_\Theta^\phi A_\psi^T  + \Sigma_Y + (\mu_Y-A_\psi \mu_\Theta^\phi) (\mu_Y-A_\psi \mu_\Theta^\phi)^T \\
&= -2  \Sigma_Y R_\phi^T A_\psi^T   + A_\psi  \Sigma_\Theta^\phi A_\psi^T  + \Sigma_Y + \mu_Y \mu_Y^T +A_\psi \mu_\Theta^\phi  \mu_\Theta^{\phi T} A_\psi^T  -A_\psi \mu_\Theta^{\phi} \mu_Y^T - \mu_Y \mu_\Theta^\phi A_\psi^T\\
&= -2  \Sigma_Y R_\phi^T A_\psi^T   + A_\psi  (\Sigma_\Theta^\phi + \mu_\Theta^\phi  \mu_\Theta^{\phi T}  )A_\psi^T  + \Sigma_Y + \mu_Y \mu_Y^T  -A_\psi \mu_\Theta^{\phi} \mu_Y^T - \mu_Y \mu_\Theta^\phi A_\psi^T\\
&= - \Sigma_Y R_\phi^T A_\psi^T    + \Sigma_Y + \mu_Y \mu_Y^T  -A_\psi \mu_\Theta^{\phi} \mu_Y^T \\
&=  \Sigma_Y (I - R_\phi^T A_\psi^T)   + (\mu_Y   -A_\psi \mu_\Theta^{\phi}) \mu_Y^T \\
\end{align*}

\subsection{Proof of VAES}\label{sec:VAESProof}

\begin{align*} \frac{\partial L_{rec}}{\partial R_\phi}  &=  -A_\psi ^\top S_\psi ^{-1}(\mu_Y-A_\psi \mu_\Theta^\phi)\mu_Y^T - A_\psi ^T S_\psi ^{-1}  \Sigma_Y  +  A_\psi ^T S_\psi ^{-1} A_\psi  R_\phi \Sigma_Y \\
\frac{\partial L_{rec}}{\partial b_\phi}  &=  -A_\psi ^\top S_\psi ^{-1}(\mu_Y-A_\psi \mu_\Theta^\phi)\\
\frac{\partial L_{rec}}{\partial Q_\phi}  &= \frac{1}{2} A_\psi ^\top S_\psi ^{-1}A_\psi\\
\frac{\partial L_{rec}}{\partial A_\psi}  &=  -S_\psi ^{-1}(\mu_Y-A_\psi \mu_\Theta^\phi) \mu_\Theta^{\phi T} - S_\psi ^{-1}  \Sigma_Y R_\phi^T +S_\psi ^{-1} A_\psi  \Sigma_\Theta^\phi \\
\frac{\partial L_{rec}}{\partial S_\psi}  &=  -\frac{1}{2}S_\psi ^{-1}(\mu_Y-A_\psi \mu_\Theta^\phi) (\mu_Y-A_\psi \mu_\Theta^\phi)^T S_\psi^{-1}  + S_\psi ^{-1}  \Sigma_Y R_\phi^T A_\psi^T  S_\psi ^{-1} - \frac{1}{2}  S_\psi ^{-1} A_\psi  \Sigma_\Theta^\phi A_\psi^T S_\psi ^{-1} -\frac{1}{2} S_\psi^{-1}  \Sigma_Y S_\psi^{-1} + \frac{1}{2} S_\psi^{-1}\\
\frac{\partial L_{reg}}{\partial R_\phi}  &=   (R_\phi \Sigma_Y R_\phi^T + Q_\phi)^{-1} R_\phi \Sigma_Y  \\
\frac{\partial L_{reg}}{\partial b_\phi}  &= 0 \\
\frac{\partial L_{reg}}{\partial Q_\phi}  &=  -\frac{1}{2} Q_\phi^{-1} + \frac{1}{2} (R_\phi \Sigma_Y R_\phi^T + Q_\phi)^{-1}\\
\frac{\partial L_{reg}}{\partial A_\psi}  &=  0\\
\frac{\partial L_{reg}}{\partial S_\psi}  &=  0
\end{align*}
The equations to be satisfied are:
\begin{align*}
A_\psi ^\top S_\psi ^{-1}(\mu_Y-A_\psi \mu_\Theta^\phi) &=0\\
- A_\psi ^T S_\psi ^{-1}  \Sigma_Y  +  A_\psi ^T S_\psi ^{-1} A_\psi  R_\phi \Sigma_Y  +(R_\phi \Sigma_Y R_\phi^T + Q_\phi)^{-1}  R_\phi \Sigma_Y &=  A_\psi ^\top S_\psi ^{-1}(\mu_Y-A_\psi \mu_\Theta^\phi) \mu_Y^T \\
A_\psi ^\top S_\psi ^{-1}A_\psi  +  (R_\phi \Sigma_Y R_\phi^T + Q_\phi)^{-1} &=  Q_\phi^{-1}\\
-S_\psi ^{-1}(\mu_Y-A_\psi \mu_\Theta^\phi) \mu_\Theta^{\phi T} - S_\psi ^{-1}  \Sigma_Y R_\phi^T +S_\psi ^{-1} A_\psi  \Sigma_\Theta^\phi  &=0 \\
S_\psi ^{-1}  \Sigma_Y R_\phi^T A_\psi^T  S_\psi ^{-1} - \frac{1}{2}  S_\psi ^{-1} A_\psi  \Sigma_\Theta^\phi A_\psi^T S_\psi ^{-1} -\frac{1}{2} S_\psi^{-1}  \Sigma_Y S_\psi^{-1}&=\frac{1}{2}S_\psi ^{-1}(\mu_Y-A_\psi \mu_\Theta^\phi) (\mu_Y-A_\psi \mu_\Theta^\phi)^T S_\psi^{-1} + \frac{1}{2}S_\psi^{-1}
\end{align*}

Consider the first equation :

If $A$ has FRR, $$R_\phi \mu_Y + b_\phi = A_\psi^+ \mu_Y$$
If $A$ has FCR,  $$R_\phi \mu_Y + b_\phi = (A_\psi^T S_\psi^{-1} A_\psi)^{-1} A_\psi^T S_\psi^{-1 } \mu_Y$$

In either case, the equation for $R_\phi$ is 
$$
[A_\psi^T S_\psi^{-1} A_\psi + (R_\phi \Sigma_Y R_\phi^T + Q_\phi)^{-1} ] R_\phi = A_\psi^T S_\psi^{-1}
$$

The equation for $Q_\phi$ is
$$A_\psi^T S_\psi^{-1} A_\psi + (R_\phi \Sigma_Y R_\phi^T + Q_\phi)^{-1} = Q_\phi^{-1}$$

Therefore,
$$R_\phi = Q_\phi A_\psi^T S_\psi^{-1}$$

Let's rewrite the equation for $Q_\phi$ using the woodbury identity

$$A_\psi^T S_\psi^{-1} A_\psi + [Q_\phi^{-1}-Q_\phi^{-1}R_\phi(\Sigma_Y^{-1}+R_\phi^T Q_\phi^{-1}R_\phi)^{-1}R_\phi^T Q_\phi^{-1}] = Q_\phi^{-1}$$

Then

\begin{align*}
A_\psi^T S_\psi^{-1} A_\psi -Q_\phi^{-1}R_\phi(\Sigma_Y^{-1}+R_\phi^T Q_\phi^{-1}R_\phi)^{-1}R_\phi^T Q_\phi^{-1} &= 0 \\
Q_\phi^{-1}R_\phi(\Sigma_Y^{-1}+R_\phi^T Q_\phi^{-1}R_\phi)^{-1}R_\phi^T Q_\phi^{-1} &= A_\psi^T S_\psi^{-1} A_\psi \\
A_\psi^T S_\psi^{-1}(\Sigma_Y^{-1}+R_\phi^T Q_\phi^{-1}R_\phi)^{-1}S_\psi^{-1} A  &= A_\psi^T S_\psi^{-1} A_\psi \\
\end{align*}

To satisfy the above equation, we need
\begin{align*}
S_\psi^{-1}(\Sigma_Y^{-1}+R_\phi^T Q_\phi^{-1}R_\phi)^{-1} &= I \\
(\Sigma_Y^{-1}+R_\phi^T Q_\phi^{-1}R_\phi) S_\psi &= I \\
R_\phi^T Q_\phi^{-1}R_\phi  &= S_\psi^{-1}  - \Sigma_Y^{-1} \\
S_\psi^{-1} A_\psi  Q_\phi A_\psi^T S_\psi^{-1}  &= S_\psi^{-1}  - \Sigma_Y^{-1} \\
A_\psi  Q_\phi A_\psi^T   &= S_\psi  - S_\psi  \Sigma_Y^{-1} S_\psi  
\end{align*}

The equations for $A_\psi$ and $S_\psi$ are the same as in VAEI. Therefore

\begin{align*}
A_\psi &= (\mu_Y \mu_\Theta^{\phi T} +  \Sigma_Y R_\phi^T)( \mu_\Theta^\phi \mu_\Theta^{\phi T} +  \Sigma_\Theta^\phi  )^{-1}  \\
S_\psi &=  \Sigma_Y (I - R_\phi^T A_\psi^T)   + (\mu_Y   -A_\psi \mu_\Theta^{\phi}) \mu_Y^T 
\end{align*}

\bibliographystyle{IEEEtran}{}
\bibliography{refs}

\end{document}